\documentclass[pra,twocolumn,preprintnumbers,amsmath,amssymb,showpacs,
nofootinbib,floatfix]{revtex4}

\usepackage{graphicx,bm,amsmath}

\makeatletter
\def\graphicscale{\twocolumn@sw{0.33}{0.4}}
\def\graphicthreescale{\twocolumn@sw{0.33}{0.4}}

\makeatother

\begin{document}

\title{Entanglement and particle correlations of Fermi gases 
in harmonic traps}

\author{Ettore Vicari }
\affiliation{Dipartimento di Fisica dell'Universit\`a di Pisa and
INFN, Pisa, Italy}

\date{\today}

\begin{abstract}
We investigate quantum correlations in the ground state of
noninteracting Fermi gases of $N$ particles trapped by an external
space-dependent harmonic potential, in any dimension.  For this
purpose, we compute one-particle correlations, particle fluctuations
and bipartite entanglement entropies of extended space regions, and
study their large-$N$ scaling behaviors.  The half-space von Neumann
entanglement entropy is computed for any dimension, obtaining $S_{\rm
HS} \approx c_l N^{(d-1)/d}\ln N$, analogously to homogenous systems,
with $c_l=1/6,\,1/(6\sqrt{2}),\,1/(6\sqrt{6})$ in one, two and three
dimensions respectively.  We show that the asymptotic large-$N$
relation $S_A \approx \pi^2 V_A/3$, between the von Neumann
entanglement entropy $S_A$ and particle variance $V_A$ of an extended
space region $A$, holds for any subsystem $A$ and in any dimension,
analogously to homogeneous noninteracting Fermi gases.
  
\end{abstract}


\maketitle

% ========================= BODY =========================
%\narrowtext

\section{Introduction}
\label{intro}

The recent developments in the experiments of dilute atom gases,
namely the achievement of Bose-Einstein condensation in dilute atomic
vapors of $^{87}$Rb and $^{23}$Na~\cite{CW-02,Ketterle-02} and the
great progress in the experimental manipulation of cold atoms in
optical lattices~\cite{BDZ-08} have provided a great opportunity to
investigate the interplay between quantum and statistical behaviors in
many-body systems.  A common feature of these experiments is the
presence of a confining harmonic potential which traps the particles
within a limited spatial region.  The capability of varying the
confining potential, which may also depend on the spatial directions,
allows also to vary the effective spatial geometry of the particle
systems, including quasi-1D geometries, see, e.g.,
Refs.~\cite{KWW-06,KWW-04,SMSKE-04,PWMMFCSHB-04,THHPRP-04,HLFSS-07}.

In this paper we investigate the quantum correlations arising within
the ground state of noninteracting Fermi gases trapped by an external
space-dependent harmonic potential.  Quantum correlations can be
characterized by the expectation values of the products of local
operators, such as the particle density and one-particle operators, or
by their integral over a space region $A$, such as the particle-number
fluctuations within $A$.  Quantum correlations are also characterized
by the fundamental phenomenon of entanglement, which gives rise to
nontrivial connections between different parts of extended quantum
systems~\cite{AFOV-08,ECP-10,rev-cc}.  A measure of entanglement is
achieved by computing von Neumann (vN) or R\'enyi entanglement
entropies of the reduced density matrix of a subsystem.  One-particle
correlations and bipartite entanglement entropies provide important
and complemetary information of the quantum behavior of many-body
systems, because they probe different features of the quantum
dynamics.

We consider Fermi gases of $N$ particles confined by harmonic traps of
arbitrary dimension, and study the large-$N$ scaling behavior of the
above-mentioned observables to characterize the quantum correlations
of the ground state. We determine the asymptotic behaviors of the
half-space entanglement entropies in any dimension, which turn out to
increase as $N^{(d-1)/d}\ln N$, analogously to homogenous
systems~\cite{CMV-12b}.  We study the relation between particle
fluctuations and entanglement entropies of extended space regions.
This is motivated by recent proposals of considering the particle
fluctuations as effective probes of many-body entanglement at zero
temperature~\cite{KRS-06,KL-09,SRL-10,SRFKL-11,SRFKLL-12,CMV-12l},
which are more easily accessible experimentally.  In homogeneous
finite-volume systems of noninteracting fermions, of any dimension,
the vN entanglement entropy $S_A$ of an extended subsystem $A$ turns
out to be closely related to the particle variance $V_A$ within
$A$. Indeed, asymptotically for a large number of particles $N$,
$S_A/V_A\approx \pi^2/3$ for any subsystem $A$ and in any
dimension~\cite{CMV-12l}, with $O(1/\ln N)$ corrections.  We show that
this asymptotic behavior also holds in the presence of a
space-dependent harmonic potential, such as the one which
characterizes recent experiments with cold atoms.  For this purpose we
present several analytical results in the large-$N$ limit, and
numerical (practically exact) results at fixed $N$ by computations
from the ground-state wave function.

The paper is organized as follows. Sec.~\ref{genrel} reports some
general expressions for the ground-state many-body wave function of
free fermion gases in a harmonic trap, and define the observables that
we consider.  Sec.~\ref{onedsy} focuses on one-dimensional (1D)
systems.  Systems in higher dimensions are considered in
Sec.~\ref{hdsy}.  Finally, in Sec.~\ref{conclu} we summarize our main
results and draw our conclusions.

\section{Ground state and observables of trapped free fermion gases}
\label{genrel}

\subsection{The ground state in a harmonic trap}
\label{gstate}

We consider a gas of $N$ noninteracting spinless fermionic particles
of mass $m$ confined within a limited space region by an external
potential.  In the following we set $\hslash=1$ and $m=1$.  The
ground-state wave function is
\begin{equation}
\Psi({\bf x}_1,...,{\bf x}_N) = {1\over \sqrt{N!}} {\rm det} 
[\psi_i({\bf x}_j)],
\label{fpsi}
\end{equation}
where $\psi_i$ are the lowest $N$ eigensolutions of the one-particle
Schr\"odinger equation
\begin{eqnarray}
H\psi_i=E_i \psi_i,\qquad H = {{\bf p}^{\,2}\over 2} + V({\bf x}).
\label{hpep}
\end{eqnarray}
A generic power-law rotational-invariant potential such as
\begin{equation}
V({\bf x}) = {1\over p}   \left(\frac{{\bf x}^{\,2}}{l^2}\right)^{p/2},
\label{vxri}
\end{equation}
where $l$ is the {\em trap size},  gives rise to a trap length scale
$\xi$ which behaves as a nontrivial power of $l$,
\begin{equation}
\xi\equiv l^{\theta}, \qquad \theta={p\over p+2},
\label{xitheta}
\end{equation}
where $\theta$ is the trap exponent~\cite{CV-10}, which does
not depend on the spatial dimension in free fermion gases.  The power
$p=2$ describes the harmonic trap, where $\xi$ is the so-called
oscillator length.  In the limit $p\to\infty$ the
system becomes equivalent to a Fermi gas confined by a hard-wall
spherical trap of radius $l$.

The one-particle energy spectrum in harmonic traps is discrete.  The
eigensolutions can be written as a product of eigenfunctions of
corresponding 1D Sch\"rodinger problems, i.e.
\begin{eqnarray}
&&\psi_{n_1,n_2,...,n_d}({\bf x}) = \prod_{i=1}^d \phi_{n_i}(x_i),
\label{prodfunc}\\
&&E_{n_1,n_2,...,n_d}= \sum_{i=1}^d e_{n_i},
\label{sunei}
\end{eqnarray}
where the $n_i$ label the eigenfunctions along the $d$ directions,
which are
\begin{eqnarray}
&&\phi_n(x) = \xi^{-1/2}{H_{n-1}(X)\over
 \pi^{1/4} 2^{(n-1)/2} (n-1)!^{1/2}} \, e^{-X^2/2}, 
 %\quad X=x/\xi,
 \label{1deigf}\\
&&e_{n} =  \xi^{-2} (n - 1/2), \quad n=1,2,... \label{Ekphih}
\end{eqnarray}
where $X=x/\xi$, and $H_n(x)$ are the Hermite polynomials.  Note
however that, although the spatial dependence of the one-particle
eigenfunctions is decoupled along the various directions, fermion
gases in different dimensions present notable differences due to the
nontrivial filling of the lowest $N$ states which provides the ground
state of the $N$-particle system.

The above one-particle eigensolutions allow us to reconstruct the
corresponding fermion-gas ground state (\ref{fpsi}), and study its
general properties by computing particle correlations and bipartite
entanglement entropies.  In the following we also set $l=1$, thus
\begin{equation}
\xi\equiv l^{\theta}=1,\quad X\equiv x/\xi=x.
\label{xXrel}
\end{equation}
The dependence on $\hslash$, $m$ and $l=\omega^{-1}$ of the quantities
considered can be easily reconstructed by a dimensional analysis.

\subsection{Observables}
\label{pce}

\subsubsection{One-particle and density correlations}
\label{pde}

The one-particle correlation function reads
\begin{eqnarray}
&&C({\bf x},{\bf x}) \equiv  
\langle c^\dagger({\bf x}) 
c({\bf y}) \rangle = 
\sum_{i=1}^N \psi_i({\bf x})^*\psi_i({\bf y})
 \label{rhonbos}
\end{eqnarray}
where $c({\bf x})$ is the fermionic annihilation operator. The
particle density and the connected density-density correlation are
respectively given by
\begin{eqnarray}
&&\rho({\bf x})\equiv 
\langle n({\bf x}) \rangle = C({\bf x},{\bf x}) =  
\sum_{i=1}^N |\psi_i({\bf x})|^2,
\label{dnbos}\\
&&G_n({\bf x},{\bf y}) \equiv 
\langle n({{\bf x}}) n({{\bf y}}) \rangle_c=\nonumber\\
&&\quad=-|C({\bf x},{\bf y})|^2 + \delta({\bf x}-{\bf y}) C({\bf x},{\bf y})
\label{gnbos}
\end{eqnarray}
where $n({\bf x})=c({\bf x})^\dagger c({\bf x})$ is the
particle-density operator, and we used the Wick theorem to write $G_n$
in terms of the two-point function $C$.

\subsubsection{Particle fluctuations and entanglement entropies}
\label{pde2}

Other important measures of the quantum correlations are related to
extended spatial regions, such as the distribution of the particle
number and the entanglement with the rest of the system.

The expectation value and connected correlators 
\begin{eqnarray}
N_A = \langle \hat{N}_A \rangle,
\quad \langle \hat{N}_A^m \rangle_c = 
\int_A \prod_{i=1}^m d^dx_i \langle 
\prod_{i=1}^m n({\bf x}_i)\rangle_c,
\label{nadef}
\end{eqnarray}
of  the  particle-number operator of an extended region $A$,
\begin{equation}
\hat{N}_A = \int_A d^dx \,n({\bf x}),
\label{hatna}
\end{equation}
characterize the particle distribution within $A$.  For this purpose,
it is convenient to introduce the cumulants of the particle
distribution, which can be defined through a generator function
as~\cite{cumgen}
\begin{equation}
V_A^{(m)}=(-i\partial_\lambda)^m
\ln \langle e^{i\lambda \hat{N}_A} \rangle |_{\lambda=0}.
\label{cumdef}
\end{equation}
In particular, the particle variance reads
\begin{equation}
V_A\equiv V_A^{(2)}=\langle N_A^2 \rangle_c\equiv
\langle N_A^2 \rangle -\langle N_A \rangle^2
\label{v2na}
\end{equation}
(the superscript $m=2$ will be understood in the case of the particle
variance).  A measure of the entanglement of the extended region $A$
with the rest of the system is provided by the 
R\'enyi entanglement entropies, defined as
\begin{equation}
S^{(\alpha)}_A = \frac{1}{1-\alpha} \ln {\rm Tr}\rho_A^\alpha
\label{saldef}
\end{equation}
where $\rho_A$ is the reduced density matrix $\rho_A$ of the subsystem
$A$. For $\alpha\to 1$, we recover the vN definition
\begin{equation}
S_A \equiv S^{(1)}_A \equiv -{\rm Tr}\,{\rho_A\ln\rho_A}
\label{criticalent}
\end{equation}
(the superscript $\alpha=1$ will be understood in the case of the
vN entanglement entropy).

In noninteracting Fermi gases the particle cumulants and the
entanglement entropies of a subsystem $A$ can be related to the
two-point function $C(x,y)$ restricted within $A$, which we denote by
${\mathbb C}_A(x,y)$. The particle number and cumulants within $A$ can
be derived using the relations (see e.g. Ref.~\cite{SRFKLL-12})
\begin{eqnarray}
&& N_A = {\rm Tr}\,{\mathbb C}_A, \label{naomc}\\
&&V_A^{(m)} = (-i\partial_z)^m {\cal G}(z,{\mathbb C}_A)|_{z=0},\label{vnyc}\\
&&{\cal G}(z,{\mathbb X}) = {\rm Tr}\ln\left[1 + \left(e^{iz} - 1\right)
{\mathbb X}\right].
\label{ygenc}
\end{eqnarray}
The vN and R\'enyi entanglement entropies can be evaluated from the
eigevalues of ${\mathbb C}_A(x,y)$ (see Refs.~\cite{JK-04,PE-09} for
applications to lattice systems).  

The computation of  particle
cumulants and entanglement entropies in 
Fermi gases of $N$ particles is much simplified by
introducing the $N\times N$ {\em overlap} matrix ${\mathbb
A}$~\cite{CMV-11,Klich-06},
\begin{equation}
{\mathbb A}_{nm} =  \int_A d^d z\, \psi_n^*(z) \psi_m(z),
\qquad n,m=1,...,N,
\label{aiodef}
\end{equation}
where the integration is over the spatial region $A$, and involves the
lowest $N$ energy levels. The overlap matrix ${\mathbb A}$ and the
restricted two-point function ${\mathbb C}_A$ satisfy
\begin{equation}
{\rm Tr} \,{\mathbb C}_A^k = {\rm Tr} {\mathbb A}^k \quad \forall 
\;k\in {\mathbb N},
\label{trca}
\end{equation}
which implies that the particle cumulants and the entanglement
entropies can be computed form the eigenvalues of the $N\times N$
overlap matrix ${\mathbb A}$.  The eigenvalues $a_i$ of ${\mathbb A}$
are real and limited, $a_i \in (0,1)$. 

The particle number and cumulants can be computed using the
relations~\cite{CMV-12l}
$N_A = {\rm Tr} {\mathbb A}$ and 
\begin{eqnarray}
&&V_A^{(m)} = (-i\partial_z)^m {\cal G}(z,{\mathbb A})|_{z=0}.\label{vny}
\end{eqnarray}
In particular,
\begin{eqnarray}
&&V_A = {\rm Tr} {\mathbb A} ( 1 - {\mathbb A}), \label{v2om}\\
&&V^{(3)}_A = {\rm Tr} [{\mathbb A}  - 3 {\mathbb A} ^2 + 2{\mathbb A}^3],  
\label{v3om}\\
&&V^{(4)}_A = {\rm Tr} [{\mathbb A} - 7 {\mathbb A}^2 + 12 {\mathbb A} ^3
- 6 {\mathbb A} ^4] , 
\label{v4om}
\end{eqnarray}
etc....  The vN and R\'enyi entanglement entropies are obtained
 by~\cite{CMV-11,CMV-11a}
\begin{equation}
S^{(\alpha)}_A = \sum_{n=1}^N s_\alpha(a_n),
\label{snx2n}
\end{equation}
where $a_n$ are the eigenvalues of ${\mathbb A}$, and
\begin{equation}
s_\alpha(\lambda) = {1\over 1-\alpha} \ln \left[{\lambda}^\alpha
+\left({1-\lambda}\right)^\alpha\right].
\label{enx}
\end{equation}
and, in particular, 
\begin{equation}
s_1(\lambda) = - \lambda \ln \lambda - (1-\lambda)\ln(1-\lambda)
\label{e1func}
\end{equation}
for the vN entropy.  We also mention that, while ${\rm Tr} {\mathbb
A}$ gives the average particle number $N_A$ within $A$, ${\rm det}{\mathbb
A}$ is the probability to find all particles within $A$.

We consider  two different partitions of the space:

(i) The subsystem $B$ is separated from the rest by a hyperplane at a
distance $x$ from the center of the trap. The corresponding
$x$-dependent particle cumulants and entanglement entropies are
denoted by $V_{B}^{(m)}(x)$ and $S_{B}^{(\alpha)}(x)$.  In one
dimension, the subsystem $B$ is given by the infinite interval
$B=[-\infty,x]$.  The half-space quantities are defined as
\begin{eqnarray}
V_{{\rm HS}}^{(m)} \equiv   V_{B}^{(m)}(0), \quad
S_{{\rm HS}}^{(\alpha)} \equiv   S_{B}^{(\alpha)}(0) \label{sa1o2}.
\end{eqnarray}
We also define
\begin{eqnarray}
S_{\Delta}^{(\alpha)}(x) \equiv 
S_{B}^{(\alpha)}(x) - S_{B}^{(\alpha)}(0) .\label{sx}
\end{eqnarray}

(ii) The subsystem $S$ is a region containing the center of trap,
and enclosed by two parallel hyperplanes at distance $x$ from the
center. The corresponding entanglement entropies are denoted by
$V_{S}^{(m)}(x)$ and $S_{S}^{(\alpha)}(x)$.  In one dimension, the
subsystem $S$ is given by the symmetric interval $S=[-x,x]$ (where $x=0$
corresponds to the center of the trap).

\section{Fermi gases in 1D traps}
\label{onedsy}

In this section we consider 1D noninteracting
spinless fermion gases of $N$ particles confined by a power-law
external potential, in particular by a harmonic potential.  This model
has a wider application, because 1D Bose gases in the limit of strong
short-ranged repulsive interactions can be mapped into a spinless
fermion gas.  The basic model to describe the many-body features of a
boson gas confined to an effective 1D geometry is the Lieb-Liniger
model with an effective two-particle repulsive contact
interaction~\cite{LL-63},
\begin{equation}
{\cal H}_{\rm LL} = \sum_{i=1}^N \left[ {p_i^2\over 2m} + V(x_i)\right] +
g \sum_{i\ne j} \delta(x_i-x_j)
\nonumber
\end{equation}
where $N$ is the number of particles and $V(x)$ is the confining
potential.  The limit of infinitely strong repulsive interactions
corresponds to a 1D gas of impenetrable bosons~\cite{Girardeau-60},
the Tonks-Girardeau gas.  1D Bose gases with repulsive two-particle
short-ranged interactions become more and more nonideal with
decreasing the particle density, acquiring fermion-like properties, so
that the 1D gas of impenetrable bosons is expected to provide an
effective description of the low-density regime of confined 1D bosonic
gases~\cite{PSW-00}.  Therefore, due to the mapping between 1D gases
of impenetrable bosons and spinless fermions, some correlations in
free fermion gases are identical to those of the hard-core boson
gases, such as those related to the particle density, particle
fluctuations of extended regions, and bipartite entanglement entropies
of connected parts.
This correspondence holds also in the presence of an external
space-dependent potential.

\subsection{The particle correlators}
\label{paco}

In 1D noninteracting Fermi systems 
with $N$ particles in a harmonic trap,
the two-point correlation function (\ref{rhonbos}) can be written as
\begin{eqnarray}
C(x,y)= {N^{1/2}\over \sqrt{2}}\,
{\phi_{N+1}(x)\phi_N(y) - \phi_{N}(x)\phi_{N+1}(y)
\over x-y},
\label{gxyha}
\end{eqnarray}
where we used
the Christoffel-Darboux relation for othornormal polynomials.
The particle density $\rho(x) = C(x,x)$,
\begin{eqnarray}
\rho(x) 
= {N^{1/2}\over \sqrt{2}}\,
\left[ \phi^\prime_{N+1}(x)\phi_N(x) - \phi^\prime_{N}(x)\phi_{N+1}(x)\right],
\label{rxyha} 
\end{eqnarray}
shows a peculiar behavior characterized by $N$ local maxima,
which get suppressed by powers of $1/N$ with increasing $N$.

Since the particle density and the density correlator of free fermion
gases are equal to those of boson gases in the hard-core limit, the
results already obtained for system of impenetrable bosons in a
trapping potential apply also to trapped fermion gases.  The large-$N$
asymptotic expansion is known to $O(1/N)$~\cite{KB-02,GFF-05}.  The
leading behavior is given by
\begin{eqnarray}
\rho(x) = N^{1/2} \left[ R_\rho(\zeta)  + O(1/N) \right],
\quad \zeta \equiv  x/N^{1/2},
\label{dnto1on} 
\end{eqnarray}
with 
\begin{equation}
R_\rho(\zeta) = {1\over \pi} \sqrt{2 - \zeta^2}, 
\qquad \zeta\le\zeta_c=\sqrt{2},
\label{ry}
\end{equation}
and $R_\rho(\zeta)=0$ for $\zeta>\zeta_c=\sqrt{2}$.

\begin{figure}[tbp]
\includegraphics*[scale=\graphicscale]{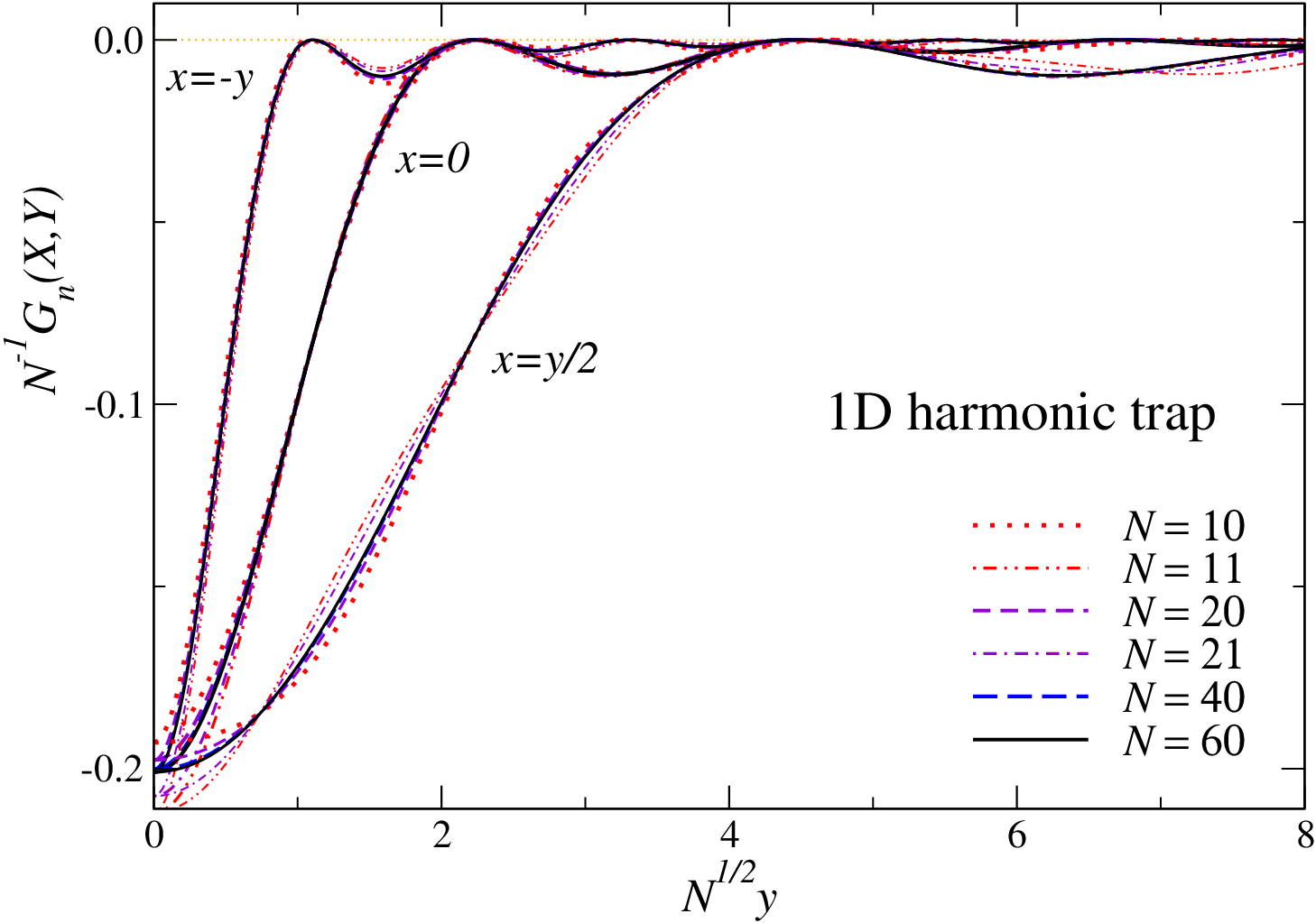}
%=% \vskip-5mm
\caption{ (Color online) Large-$N$ scaling of $G_n(x,y)$ for $x=y/2$,
$x=0$, and $x=-y$ for 1D fermion gases in a harmonic trap.
We plot $N^{-1}G_n(x,y)$ vs $N^{1/2}y$.  }
\label{gnxp2ln}
\end{figure}

The space dependence of the connected correlation function of the
particle density operator $n_{x}=c(x)^\dagger c(x)$ presents a
different large-$N$ scaling behavior, characterized by different power
laws~\cite{CV-10-bhn}, i.e.,
\begin{equation}
G(x,y) 
\approx N R_G(N^{1/2}x,N^{1/2} y),
\label{gnbosln}
\end{equation}
for $x\ne y$, as shown by Fig. ~\ref{gnxp2ln}.  Note that the
asymptotic regime of $G_n$ is not approached uniformly when $x\to y$,
because $G(x,x)=\rho(x)-\rho(x)^2$, cf. Eq.~(\ref{gnbos}).

The large-$N$ scaling function of the particle density in a harmonic
trap vanishes at $\zeta_c=\sqrt{2}$ where $R_\rho(\zeta_c)=0$.
Around this point the particle correlations~\cite{CV-10-bhn}, and also the
entanglement entropies, show a different large-$N$ scaling behaviors
characterized by other power laws~\cite{CV-10-e}.  Indeed, around the point
$x_c\equiv N^{1/2}\zeta_c$, the particle density and its correlation
behave as
\begin{eqnarray}
&&\rho(x) \approx N^{1/6} g_\rho[N^{1/6}(x-x_c)],\quad
x_c=N^{1/2}\zeta_c,
\quad\label{rhoxbo}\\
&&G_n(x_c,x) \approx N^{1/3} g_n[N^{1/6}(x-x_c)].
\label{gxcx}
\end{eqnarray}
The scaling function $g_\rho(z)$ can be obtained from related
computations within the Gaussian unitary ensembles of random
matrices~\cite{Forrester-93,GFF-05}:
\begin{eqnarray}
g_\rho(z) &=& {\rm Lim}_{N\to\infty}
N^{-1/6}\rho[N^{1/2}(\zeta_c + N^{-2/3}z)]  \nonumber\\
&=&2^{1/2} |{\rm Ai}^\prime(2^{1/2}z)|^2 -  2z|{\rm Ai}(2^{1/2}z)|^2 .
\label{fez}
\end{eqnarray}

\subsection{Spatial entanglement}
\label{1dse}

\subsubsection{Half-space entanglement entropy}
\label{1dsea}

The asymptotic large-$N$ behavior of the half-space ($A=[-\infty,0]$
where $x=0$ is the center of the trap) vN and R\'enyi entanglement
entropies can be inferred by exploiting known results for the 1D
hard-core Bose-Hubbard model in the presence of an external power-law
potential $V(x)=(x/l)^p$ 
and a chemical potential, which is equivalent to a lattice free-fermion
model.  The derivation is outlined in App.~\ref{bentBH}.
We obtain
\begin{eqnarray}
&&S_{{\rm HS}}^{(\alpha)} =  S_{{\rm ASY}}^{(\alpha)}  + o(N^0),
\label{asyt1o2}\\
&&S_{{\rm ASY}}^{(\alpha)} = 
C_\alpha
\left[ \ln N + \ln {4(p+2)\over p} + y_\alpha\right],
\label{sasya}
\end{eqnarray}
where
\begin{eqnarray}
C_\alpha = {1+\alpha^{-1}\over 12}, 
\label{calpha}
\end{eqnarray}
$p$ is the power-law of the potential, and $y_\alpha$ is given in
Eq.~(\ref{yalpha}).  Note that the leading logarithmic term, and in
particular its coefficient, is independent of the trapping potential,
and it is equal to that of homogeneous systems with open boundary
conditions~\cite{CMV-11,CMV-11a}, which is determined by the
corresponding conformal field theory~\cite{CC-04} with central charge
$c=1$.  The asymptotic behavior (\ref{sasya}) in the limit
$p\to\infty$ reproduces the results for homogeneous systems with open
boundary conditions~\cite{CMV-11,CMV-11a}
\begin{equation}
S_{{\rm HS}}^{(\alpha)}=
C_\alpha \left[ \ln N +  
\ln 4 + y_\alpha +  O(N^{-1/\alpha})\right] ,\label{ob1o21d}
\end{equation}
because in the limit $p\to\infty$ the system becomes
equivalent to a Fermi gas confined by a 1D hard-wall trap of size
$L=2l$.

\begin{figure}[tbp]
\includegraphics*[scale=\graphicscale]{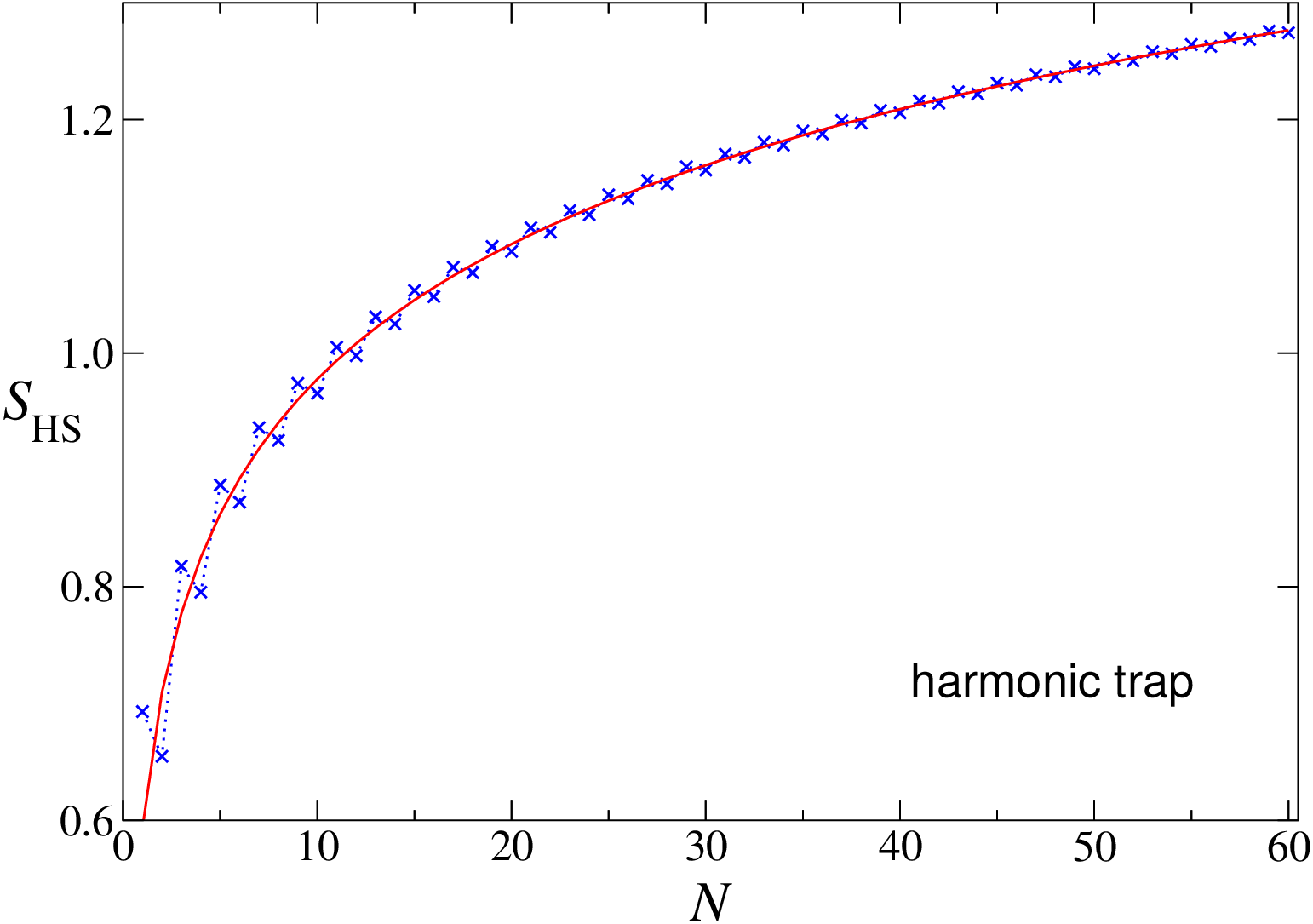}
%=% \vskip-5mm
\caption{(Color online) The half-space vN entanglement entropy.  
The full line shows the large-$N$ asymptotic behavior
(\ref{sasya}).  }
\label{hls1}
\end{figure}

\begin{figure}[tbp]
\includegraphics*[scale=\graphicscale]{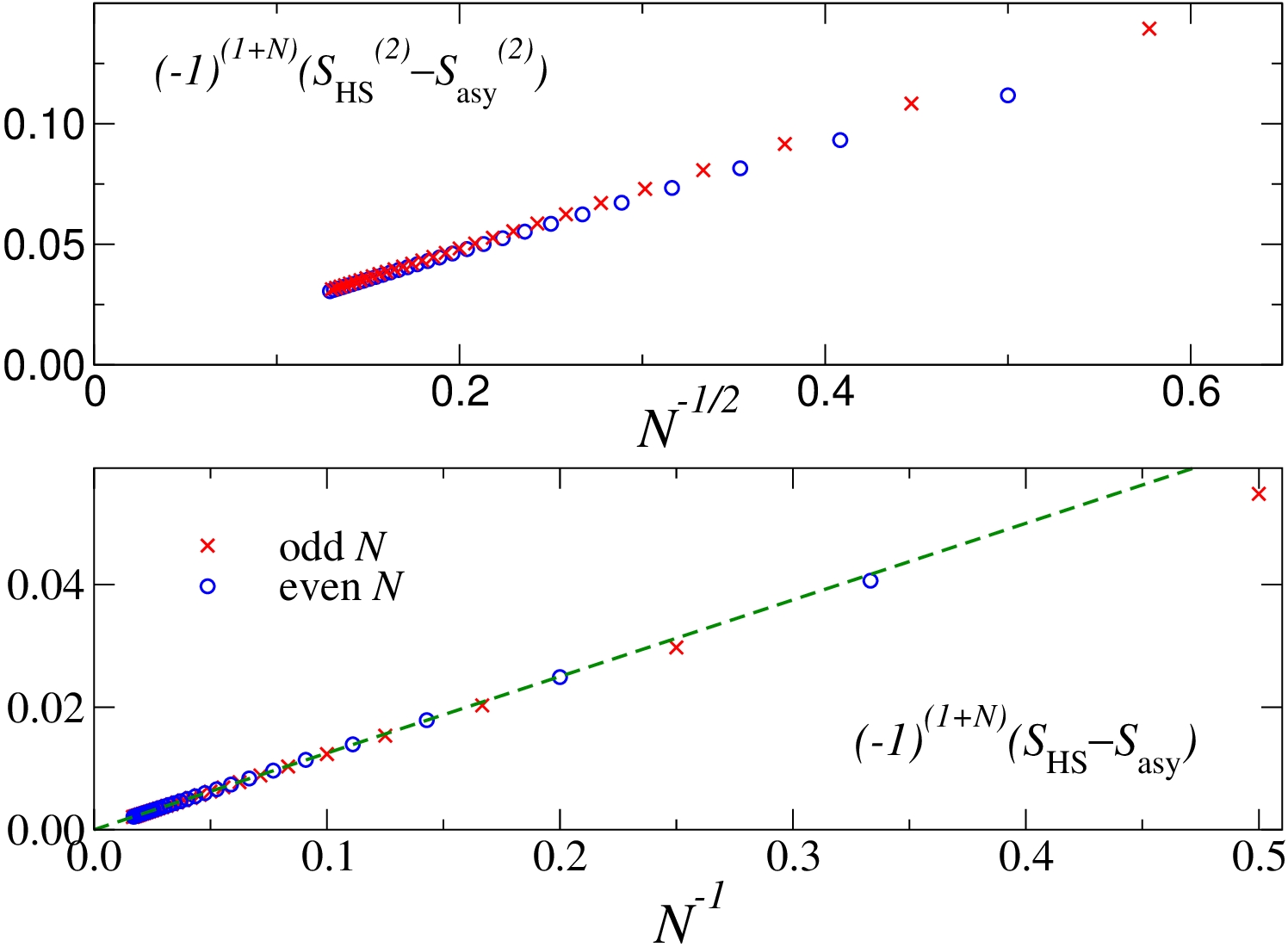}
%=% \vskip-5mm
\caption{(Color online) We plot $(-1)^{1+N}(S_{\rm HS} - S_{{\rm
ASY}})$ vs $N^{-1}$ (bottom) and $(-1)^{1+N}(S_{{\rm HS}}^{(2)} -
S_{{\rm ASY}}^{(2)})$ vs $N^{-1/2}$ (top). In the bottom figure the
dashed line shows the slope $1/8$.  }
\label{hls12c}
\end{figure}

In order to check the convergence to this asymptotic behavior, we
numerically compute the half-space entanglement entropies $S_{{\rm
HS}}^{(\alpha)}$ of $N$ particles in the presence of a harmonic trap.
We use the method based on the overlap matrix (\ref{aiodef}), i.e. we
numerically compute its eigenvalues and then obtain the entanglement entropies
through Eq.~(\ref{snx2n}).  Figs.~\ref{hls1} and \ref{hls12c} show
data for the vN and
$\alpha=2$ R\'enyi entropies. They 
are fully consistent with the asymptotic behavior (\ref{sasya}). 
In particular, the large-$N$ behavior of the vN entropy
turns out to behave as
\begin{equation}
S_{{\rm HS}} = S_{{\rm ASY}} + 
(-1)^{N} {c_1\over N}  + 
{c_{2}\over N^2}  + (-1)^{N} {c_{3}\over N^3}  + ...
\label{snalphanw}
\end{equation}
with $c_1\approx -1/8$ (with a precision better than $10^{-6}$, see
Fig.~\ref{hls12c}, $c_{2}\approx 0.009893$ and $c_{3}\approx 0.066$
(the uncertainty should be on the last figures).
The data of the $\alpha=2$ R\'enyi entropy, shown in Fig.~\ref{hls12c},
fits the Ansatz 
\begin{equation}
S_{{\rm HS}}^{(2)} = S^{(2)}_{{\rm ASY}} + 
(-1)^{N} {b_1\over N^{1/2}}  + {b_2\over N} + ...
\label{snalphanw2}
\end{equation}
with $b_1\approx -0.2387$ and $b_2\approx 0.0146$.  Note that the
above numerical results show that the corrections to the large-$N$
asymptotic behavior are $O(N^{-1/\alpha})$ in Eq.~(\ref{asyt1o2}),
analogously to homogeneous systems, cf. Eq.~(\ref{ob1o21d}).

Finally, we mention that the effects of a power-law trapping potential
on the scaling behavior of the entanglement at the quantum critical
point of 1D lattice models were investigated in
Refs.~\cite{CV-10-bh,CV-10-e,FC-08,VS-12}. In particular, the
1D hard-core Bose-Hubbard model, which is equivalent to a
free fermion lattice model, was considered in the superfluid phase at
half filling~\cite{CV-10-e}. As shown by the arguments reported in
App.~\ref{bentBH}, used to derive Eq.~(\ref{sasya}), these results are
somehow related with the large-$N$ scaling of 1D Fermi
gases investigated in this section, in particular when the chemical
potential is driven toward the superfluid-to-empty transition. 
However the large-$N$ scaling behavior of
1D continuum Fermi gases presents distinct features,
as pointed out in Refs.~\cite{CMV-11,CMV-11a} in the case
of homogenous systems.

\subsubsection{Finite intervals around the center of the trap}
\label{1dseb}

\begin{figure}[tbp]
\includegraphics*[scale=\graphicscale]{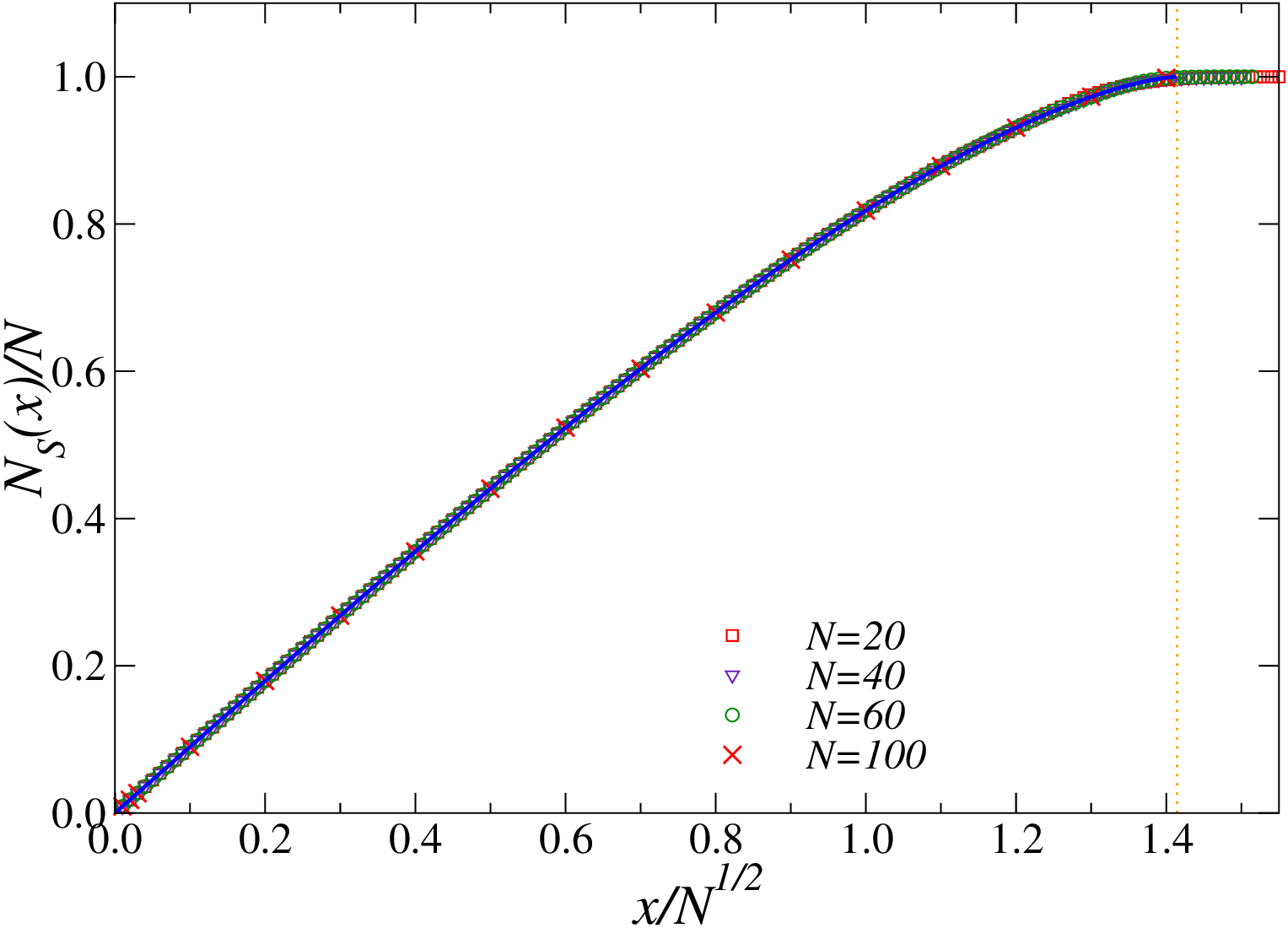}
%=% \vskip-5mm                   
\caption{ (Color online) The particle number within the interval
$S=[-x,x]$ around the center of the trap, for some values of $N$ up to
$N=100$, versus $\zeta\equiv x/N^{1/2}$.  The full line 
(hardly visible among the data symbols) shows the
large-$N$ limit (\ref{nbxn}) of the ratio $N_S(x)/N$.  }
\label{pnx}
\end{figure}

We now consider a symmetric interval $S=[-x,x]$ around the center of
the trap.  By integrating the large-$N$ particle density
(\ref{dnto1on}) within the interval $S=[-x,x]$, we obtain the average
number $N_S(x)$ of particles within $S$ in the large-$N$ limit, 
\begin{equation}
{N_S(x)\over N}
 = {1\over \pi}  \left[
 \zeta \sqrt{2-\zeta^2} + 2 {\rm arcsin}(\zeta/\sqrt{2})\right]
+O(1/N)
\label{nbxn}
\end{equation}
where $\zeta=x/N^{1/2}$.  Fig.~\ref{pnx} shows results for the
particle number within $S$ at fixed $N$, up to $N=100$. They 
show that the large-$N$ limit (\ref{nbxn}) is rapidly approached by
the data.

\begin{figure}[tbp]
\includegraphics*[scale=\graphicscale]{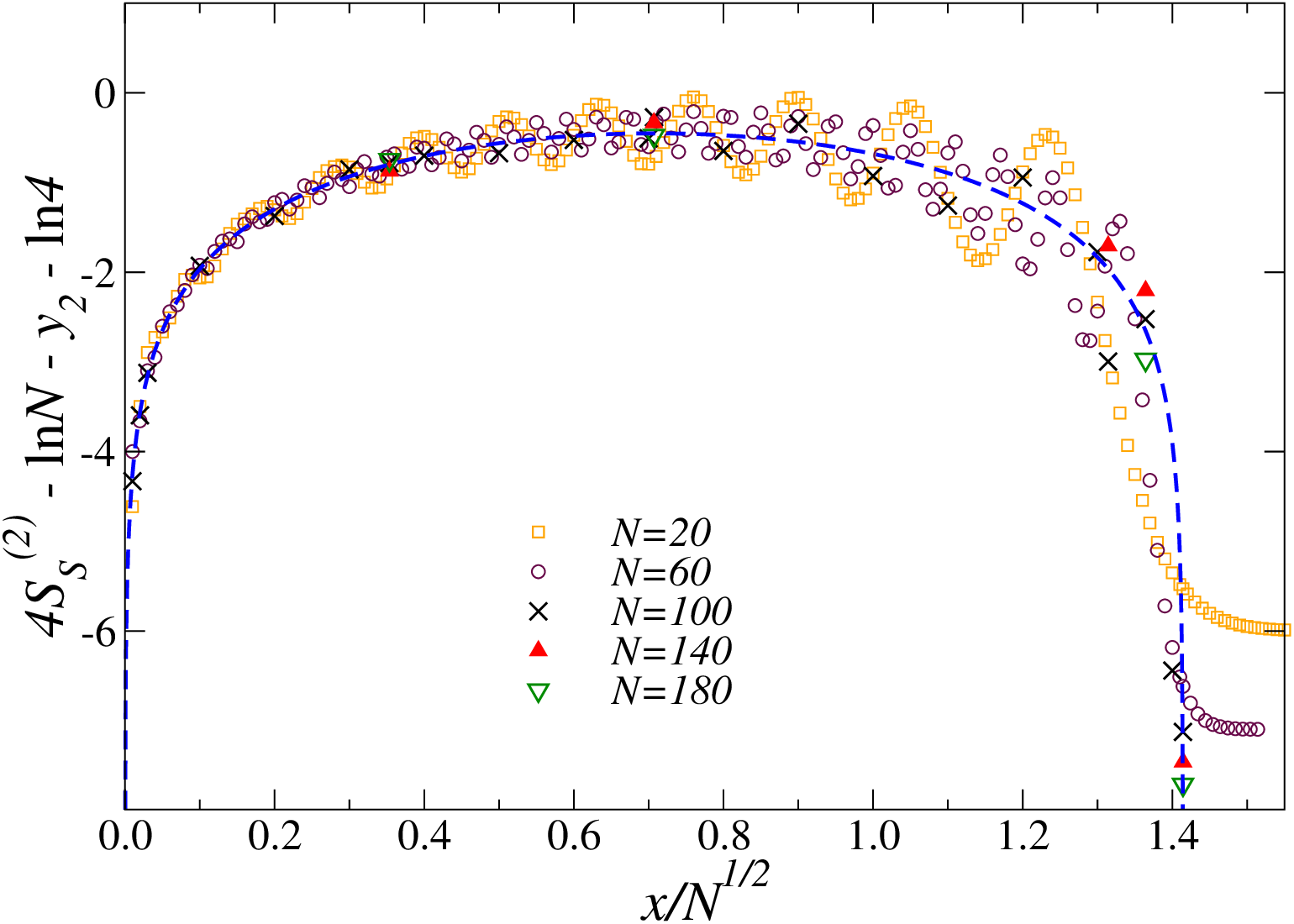}
\includegraphics*[scale=\graphicscale]{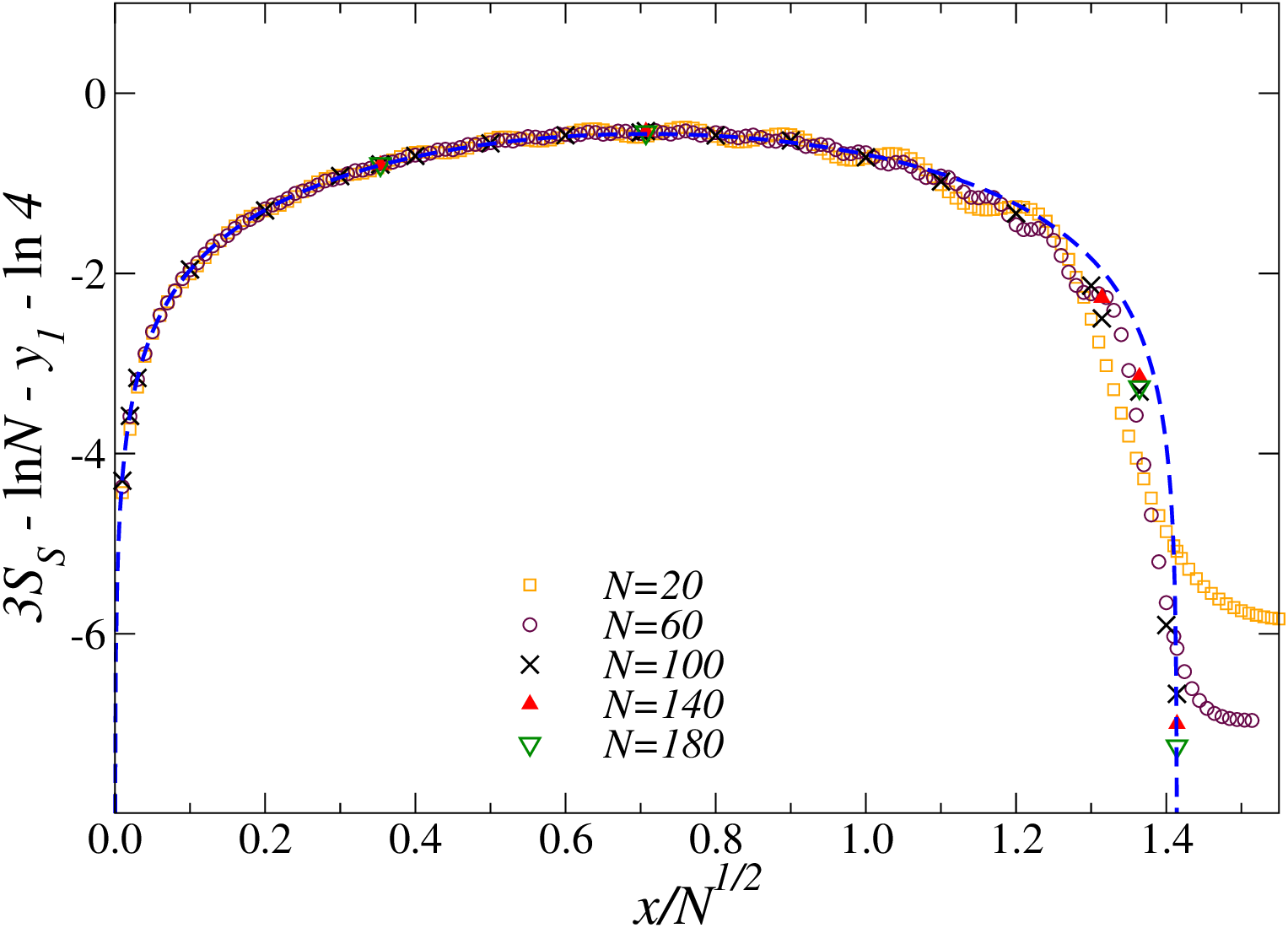}
%=% \vskip-5mm  
\caption{ (Color online) The $\alpha=1$ vN (bottom) and $\alpha=2$
R\'enyi (top) entanglement entropies of the interval $S\equiv [-x,x]$
vesus $\zeta\equiv x/N^{1/2}$.  We plot $S_S^{(\alpha)}(x)/(2 C_\alpha) -
(\ln N+y_\alpha+\ln 4)$ for some values of $N$ up to $N=180$.
The full lines show the function (\ref{fsal}).  }
\label{s1x}
\end{figure}

Results for the vN and $\alpha=2$ entanglement entropies up to $N=180$
are shown in Figs.~\ref{s1x}.  With increasing $N$, the subtracted
data of $S_S^{(\alpha)}-2C_\alpha\ln N$ appear to approach a function
of $\zeta\equiv x/N^{1/2}$.  Therefore, we infer the large-$N$ scaling
behavior
\begin{equation}
S_{S}^{(\alpha)}(x)  \approx   2C_\alpha
\left[ \ln N + y_\alpha + \ln 4 + f_S^{(\alpha)}(\zeta) \right]
\label{smxx}
\end{equation}
The scaling functions $f_{S}^{(\alpha)}(\zeta)$ are expected to be
singular at $\zeta=0$ corresponding to a vanishing interval, and at
$\zeta=\sqrt{2}$, which corresponds to the point where the particle
density vanishes in the large-$N$ limit, cf. Eq.~(\ref{dnto1on} ).
Note that the space dependence scales analogously to that of the
particle density, cf. Eq.~(\ref{dnto1on}), while it differs from that
of the connected density-density correlation, cf Eq.~(\ref{gnbosln}).

\begin{figure}[tbp]
\includegraphics*[scale=\graphicscale]{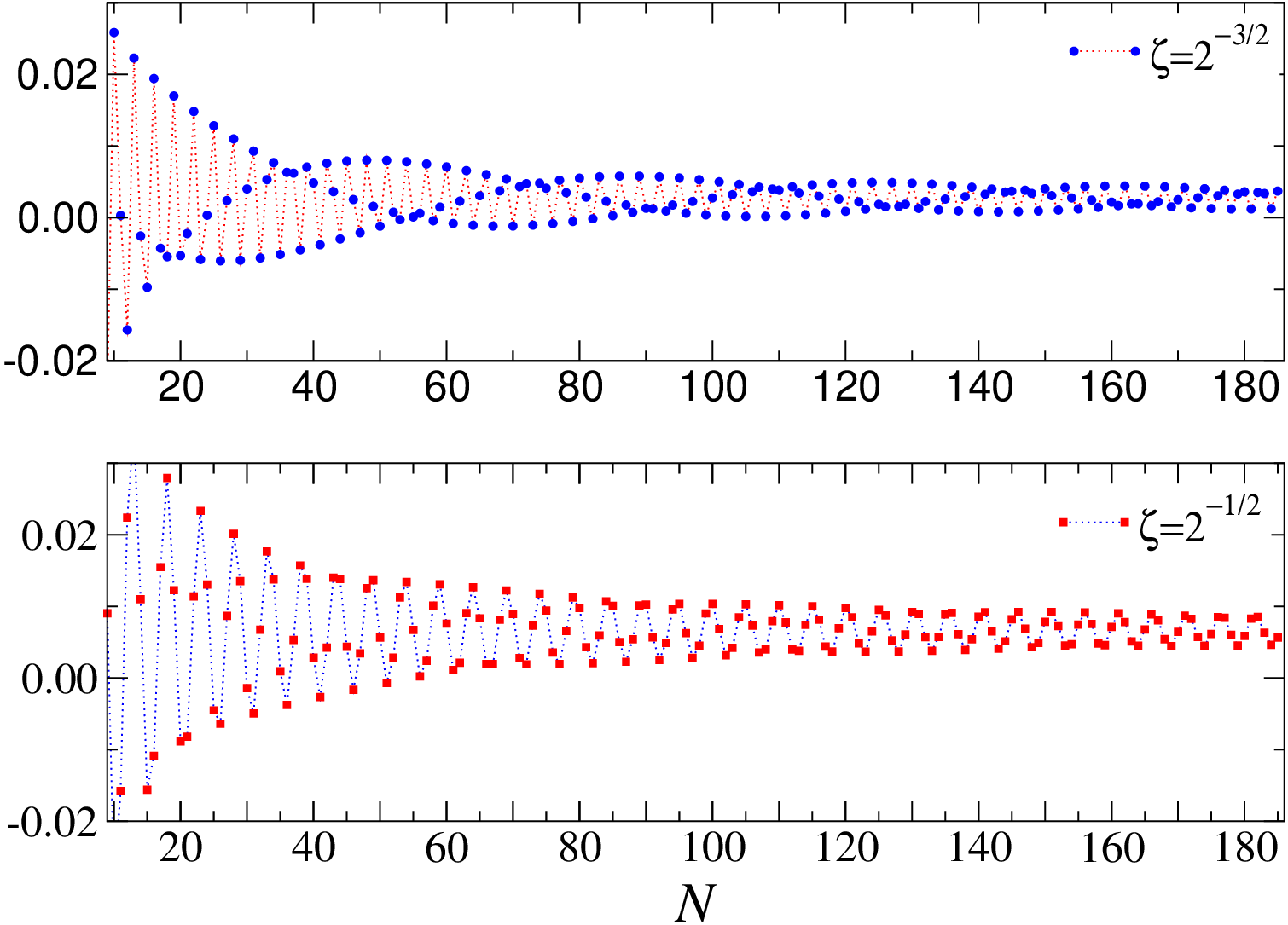}
%=% \vskip-5mm  
\caption{ (Color online) Differences between $S_S(x)$ and the
asymptotic behavior (\ref{smxx}) with $f_S$ given by Eq.~(\ref{fsal})
at $\zeta=2^{-1/2}$, which is the maximum of Eq.~(\ref{fsal}),
and $\zeta=2^{-3/2}$, versus $N$.  }
\label{s1xc}
\end{figure}

The large-$N$ convergence is rapid at least up to $\zeta\approx
1$, but also the data for $1\lesssim \zeta \lesssim \sqrt{2}$ appear
to approach a unique curve, although more slowly.  Moreover, the
behavior of the data with increasing $N$ suggests that the large-$N$
scaling functions $f_S^{(\alpha)}(\zeta)$ are independent of
$\alpha$.  Actually, they turn out to be well approximated by the
simple function
\begin{eqnarray}
f_S^{(\alpha)}(\zeta) \approx f_a(\zeta) = 
\ln\sin(\pi \zeta/\sqrt{2}) + {\rm ln}(2/\pi),
\label{fsal}
\end{eqnarray}
as shown in Fig.~\ref{s1x} for the $\alpha=1$ vN and $\alpha=2$
R\'enyi entropies. In Fig.~\ref{s1xc} we show the differences between
the vN entropy $S_S(x)$ and the asymptotic behavior (\ref{smxx}) with
$f_S$ given by Eq.~(\ref{fsal}), at its maximum $\zeta=2^{-1/2}$ and
at $\zeta=2^{-3/2}$. For example at $\zeta=2^{-1/2}$ the data show
deviations smaller than 0.01 for $N\gtrsim 100$, suggesting that the
deviation in the large-$N$ limit should be less than 0.01.  Smaller
deviations are observed at $\zeta=2^{-3/2}$, see Fig.~\ref{s1xc}.  A
more precise large-$N$ extrapolation is made difficult by the presence
of oscillations, whose structure is not clear, see Fig.~\ref{s1xc}.

It is worth comparing the above results
with the behavior of analogous quantities in
homogeneous Fermi gas within hard walls, whose entanglement
entropies of the interval $S=[-x,x]$ around the center of the
hard-wall trap of size $L=2l=2$ are given by~\cite{CMV-11a}
\begin{equation}
S_{S}^{(\alpha)}(x)  \approx   2C_\alpha
\left[ \ln N + \ln\sin(\pi x ) +                              
y_\alpha + {\rm ln}2 + O(N^{-1/\alpha})\right]
\label{smxxhw}
\end{equation}

\begin{figure}[tbp]
\includegraphics*[scale=\graphicscale]{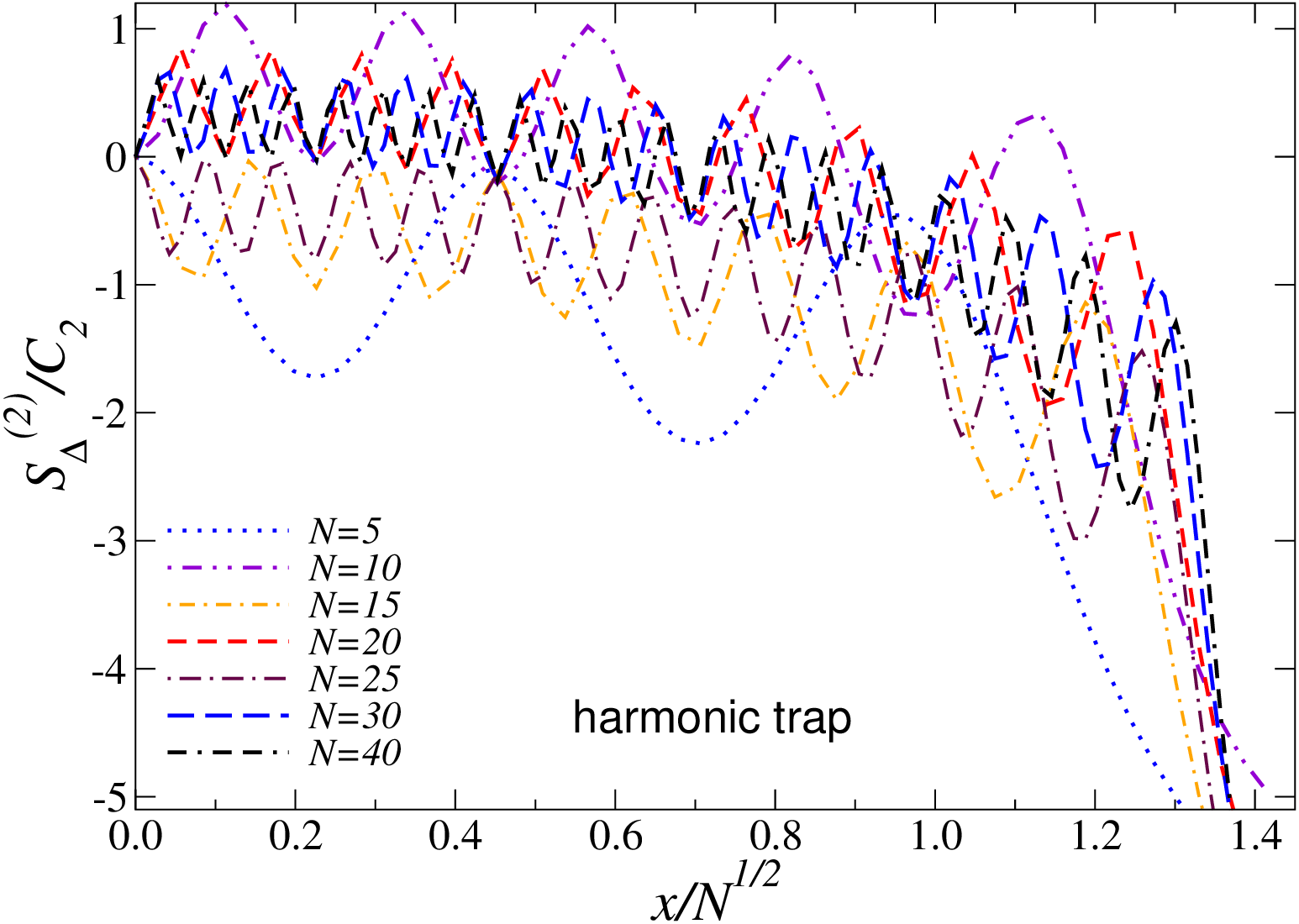}
\includegraphics*[scale=\graphicscale]{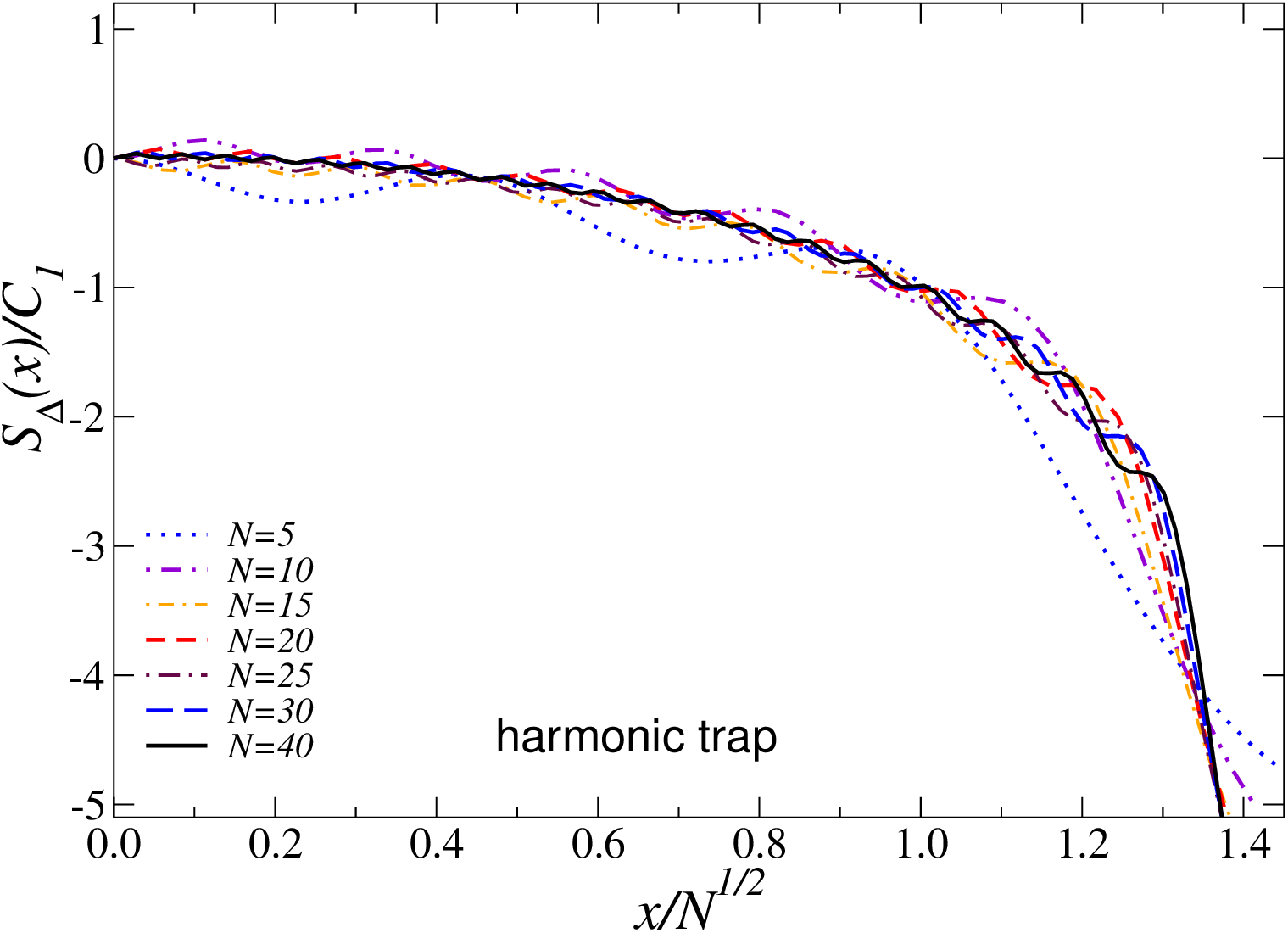}
%=% \vskip-5mm
\caption{(Color online) $S_\Delta^{(\alpha)}(x)/C_\alpha$ versus
$x/N^{1/2}$ for the $\alpha=1$ vN (bottom) and $\alpha=2$ (top)
entropies, and several values of $N$.  The two sets of data appear to
approach the same large-$N$ limit.}
\label{hls1x}
\end{figure}

Finally, Fig.~\ref{hls1x} shows results for the quantity
$S_\Delta^{\alpha}(x)\equiv S^{\alpha}_B(x) - S^{\alpha}_{\rm HS}$,
i.e. the difference between the entanglement entropies of the
intervals $[-\infty,x]$ and $[-\infty,0]$.  They show the large-$N$
scaling behavior
\begin{equation}
S^{(\alpha)}_\Delta(x)
 \approx C_\alpha f_\Delta^{(\alpha)}(\zeta),\qquad \zeta\equiv x/N^{1/2}.
\label{lnbars}
\end{equation}
They also suggest that $f_\Delta^{(1)}=f_\Delta^{(2)}$,
i.e. $f_\Delta^{(\alpha)}$ is independent of $\alpha$, although the
convergence of the $\alpha=2$ R\'enyi entropy is slower than that of the
vN entropy, apparently $O(N^{-1/2})$ against $O(N^{-1})$.

\subsection{Particle fluctuations in extended spatial subsystems}
\label{hfpf}

\begin{figure}[tbp]
\includegraphics*[scale=\graphicscale]{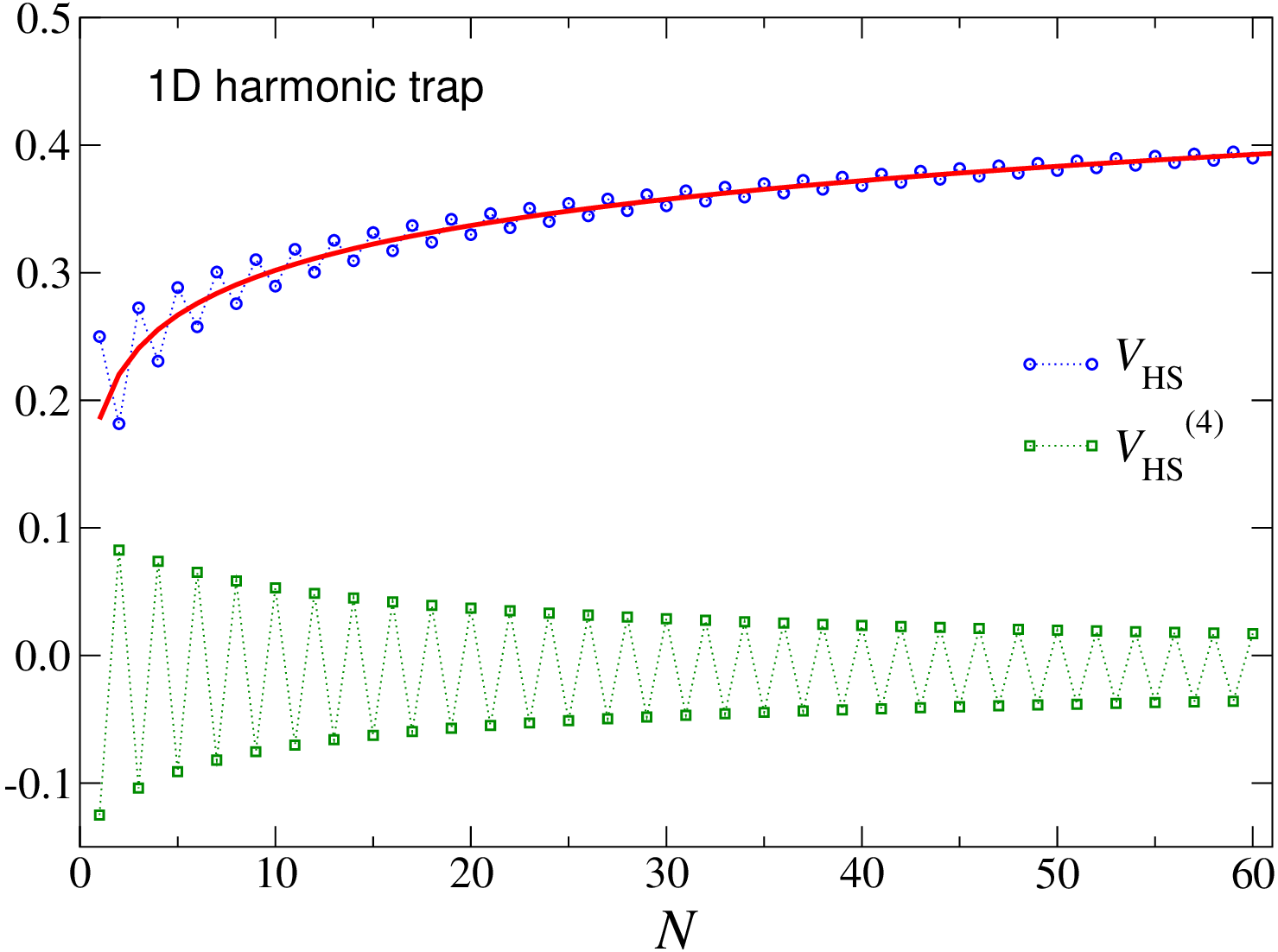}
%=% \vskip-5mm
\caption{ (Color online) The half-space particle-number cumulants 
$V_{\rm HS}$ and $V_{\rm HS}^{(4)}$. The full line shows
the function (\ref{v2hstrap}).
 }
\label{1dha}
\end{figure}

\begin{figure}[tbp]
\includegraphics*[scale=\graphicscale]{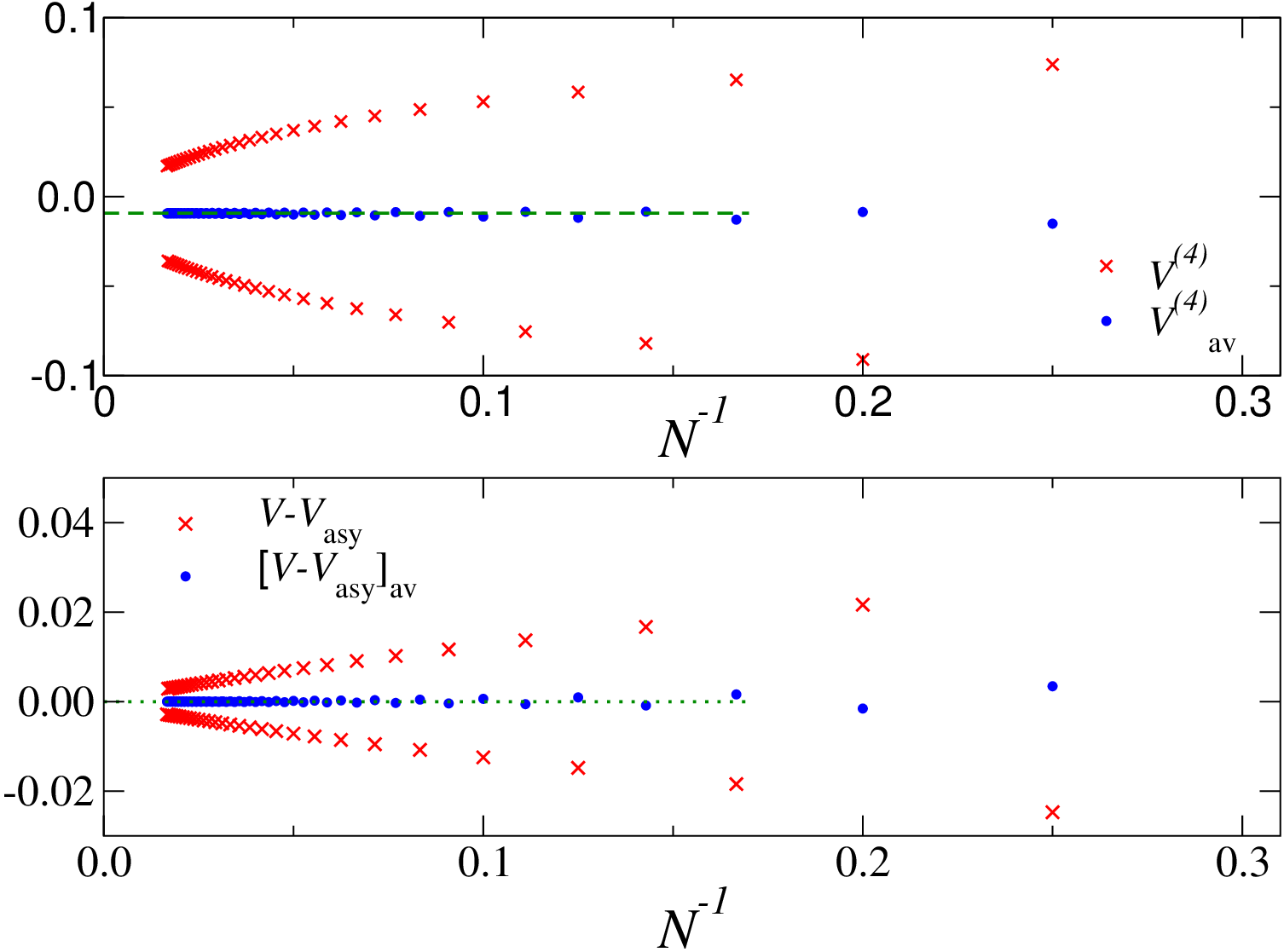}
%=% \vskip-5mm
\caption{ (Color online) Check of the asymptotic behavior for the
particle variance and quartic cumulant of half space.  The subscript
"${\rm av}$" indicates the average over the last two data to suppress
the large oscillations.  The dotted and dashed lines show the
$N\to\infty$ limit expected for $V_{\rm HS}-V_{\rm ASY} \approx 0$ and
$V_{\rm HS}^{(4)}\approx -0.009255$.  }
\label{1dhb}
\end{figure}

Some results for the half-space particle variance and quartic
cumulant are shown in Fig.~\ref{1dha}.  They are
characterized by large odd-even oscillations in the number of
particles.  An educated guess for the asymptotic large-$N$ behavior of
the half-space particle variance in a generic external potential
$V(x)\propto (x/l)^p$ is
\begin{eqnarray}
&&V_{\rm HS} \approx  V_{\rm ASY} + o(N^0),\nonumber \\
&&V_{\rm ASY}=
{1\over 2\pi^2} \left[ \ln N + \ln {4(p+2)\over p} + 1+\gamma_E  \right].
\label{v2hstrap}
\end{eqnarray}
This asymptotic behavior is somehow derived by analogy with the
asymptotic behavior of the half-space R\'enyi entanglement entropies,
taking also into account the known 
asymptotic behavior of the particle variance in hard-wall
traps~\cite{CMV-12l},  
\begin{eqnarray}
V_{\rm HS} = {1\over 2\pi^2} \left[ \ln N + 
\ln 4 + 1 + \gamma_E + O(N^{-1})\right],
\label{v2hstraphw}
\end{eqnarray} 
which must be recovered in $p\to\infty$ limit.

Concerning the other cumulants, we expect that the leading term is the
same as that of homogenous systems within hard walls, like the leading
terms of the entanglement entropies and particle variance.  Thus
\begin{eqnarray}
V^{(2i)}_{\rm HS} =  \nu_{2i} + o(N^0) \quad{\rm for}\;\;i>2\label{v2ih}
\end{eqnarray}
where $v_{2i}$ are the same constant appearing in the case of the
hard-wall trap~\cite{CMV-12l}, i.e. $\nu_4=-0.0092552$,
$\nu_6=0.00404469$, etc...

The above large-$N$ predictions are fully supported by 
the numerical data at fixed $N$ with increasing $N$, as shown in
Fig.~\ref{1dhb}, where we also show data averaged over two subsequent
particle numbers to suppress the odd-even oscillations.  In the case
of the particle variance, the amplitude of the odd-even oscillations
appear to decrease as $O(1/N)$, while the average between the data for
subsequent particle numbers approaches the predicted asymptotic
behavior much more rapidly.  In the case of the quartic cumulant, the
oscillations get suppressed more slowly, but their odd-even average
approaches the predicted value quite rapidly.  For the largest available
values of $N$ the difference of these averages from the asymptotic
predicted behaviors is $O(10^{-5})$.

\begin{figure}[tbp]
\includegraphics*[scale=\graphicscale]{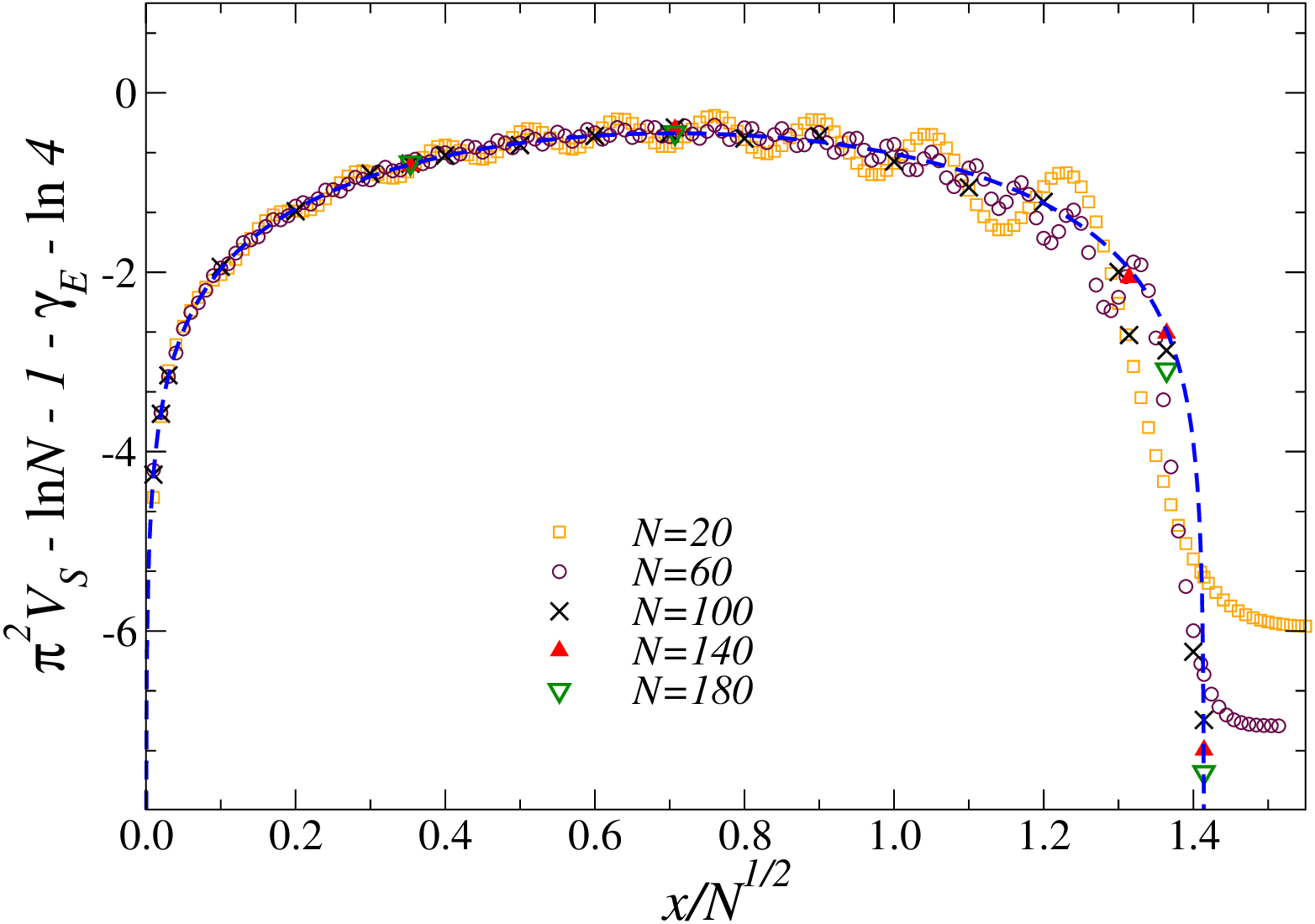}
%=% \vskip-5mm                                
\caption{ (Color online) The particle variance 
of intervals $S=[-x,x]$ for some values of $N$,
versus $\zeta=x/N^{1/2}$.  We plot $\pi^2 V - (\ln N + 1 + \gamma_E + \ln 4)$
The line shows the function (\ref{fsal}).  }
\label{v2x}
\end{figure}

\begin{figure}[tbp]
\includegraphics*[scale=\graphicscale]{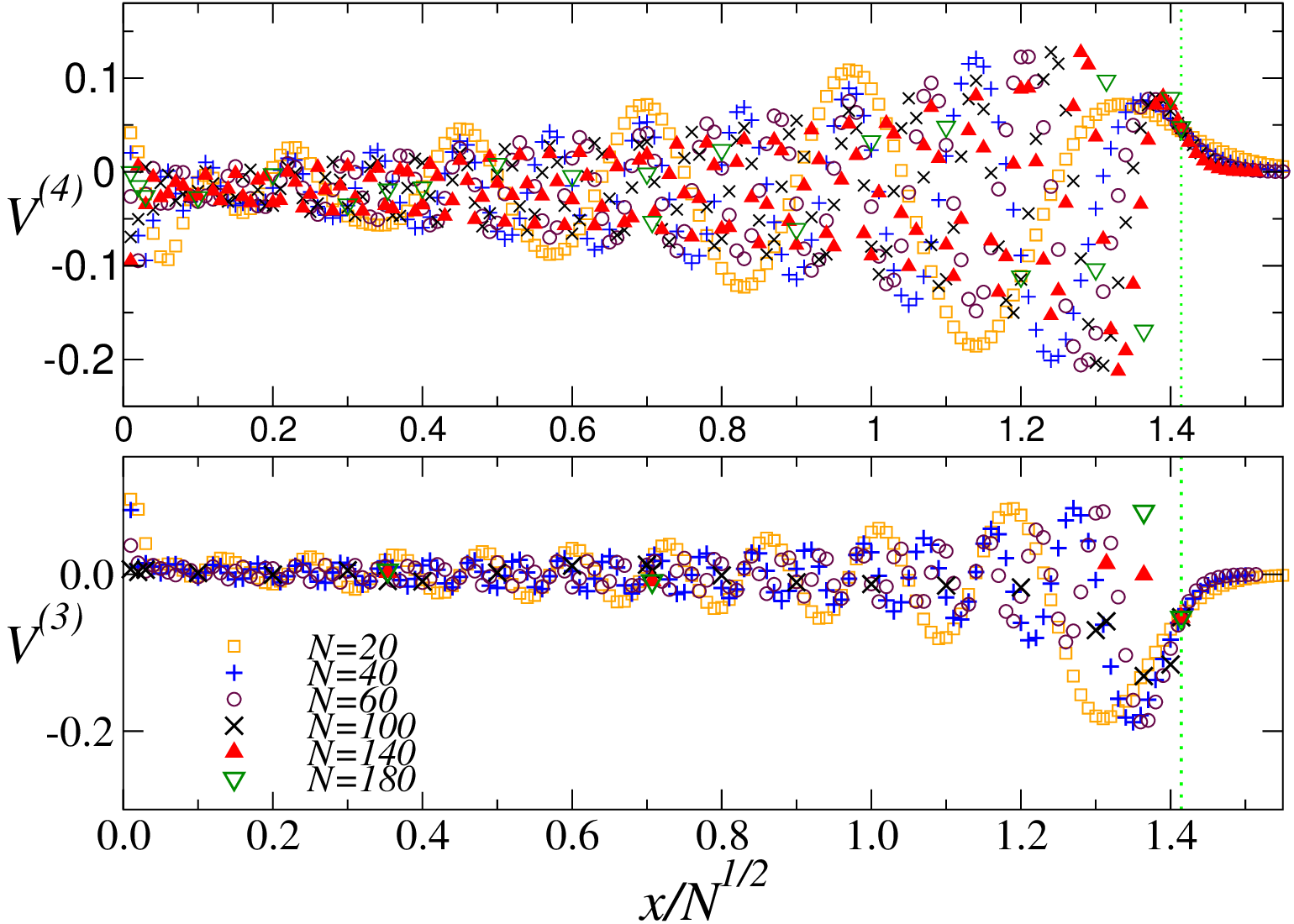}
%=% \vskip-5mm                                
\caption{ (Color online)
The third (top)  and quartic (bottom) cumulants of the interval $S=[-x,x]$.
}
\label{v34x}
\end{figure}

We now consider the interval $S=[-x,x]$ around the center of the trap.
In Fig.~\ref{v2x} we show the particle variance for values of $N$ up
to $N=180$.  They show a behavior analogous to that of the 
entanglement entropies, see Fig.~\ref{s1x}, and are consistent with 
\begin{equation}
V_S(x) \approx  {1\over \pi^2} 
\left[ \ln N + 1 + \gamma_E + \ln 4 + f_V(\zeta) \right]
\label{fsalvsx}
\end{equation}
The analysis of the data with increasing $N$ is consistent with
the relation $f_V(\zeta)=f_S^{(\alpha)}(\zeta)$, where
$f_S^{(\alpha)}(\zeta)$ are the corresponding scaling functions of the
entangelement entropies, cf. Eq.~(\ref{smxx}).  Therefore, $f_V(\zeta)$ is well
approximated by the same function $f_a(\zeta)$, cf. Eq.~(\ref{fsal}).
Again, this behavior resembles that of the homogeneous system within a
hard-wall trap of size $L=2l=2$, which is~\cite{CMV-12l}
\begin{eqnarray}
V_S(x)  &=& {1\over \pi^2}
\Bigl[ \ln N + \ln\sin(\pi x ) + \label{v2xxhw}\\           
&&+1 + \gamma_E  + {\rm ln}2 + O(N^{-1})\Bigr]
\nonumber
\end{eqnarray}
In Fig.~\ref{v34x} we show the third and quartic cumulants of the
interval $S=[-x,x]$. They are characterized by oscillations which
increase when $\zeta\to\sqrt{2}$, but remain apparently limited with
increasing $N$.

\section{Higher-dimensional systems}
\label{hdsy}

In this section we consider Fermi gases confined by two- and
three-dimensional traps. We again study the large-$N$ behavior of the
particle correlators, cumulants of the particle-number distribution
and entanglement entropies of extended spatial regions.

The vN and R\'enyi entanglement entropies of extended spatial
subsystems in the ground state of homogenous Fermi gases 
of $d$ dimension grow
asymptotically as $N^{(d-1)/d} \ln N$, with a prefactor that is
analytically computed using the Widom conjecture~\cite{Widom-81}, for
both periodic and open boundary conditions.  The logarithmic
correction to the power-law behavior is related to the area-law
violation in lattice free
fermions~\cite{Wolf-06,GK-06,BCS-06,LDYRH-06,FZ-07,HLS-09,DBYH-08,Swindle-10},
i.e.  for a large subsystem $A$ of linear size $\ell$ in an infinite
$d$-dimensional lattice the entanglement entropies scale like
$S^{(\alpha)}(A) \sim \ell^{d-1} \ln \ell$.  In this section we study
the effects of a space-dependent confining potential in 2D and 3D 
Fermi systems, investigating again the relations
between particle fluctuations and entanglement entropies.

\subsection{Particle density and its correlator}
\label{2dpdc}

\begin{figure}[tbp]
\includegraphics*[scale=\graphicscale]{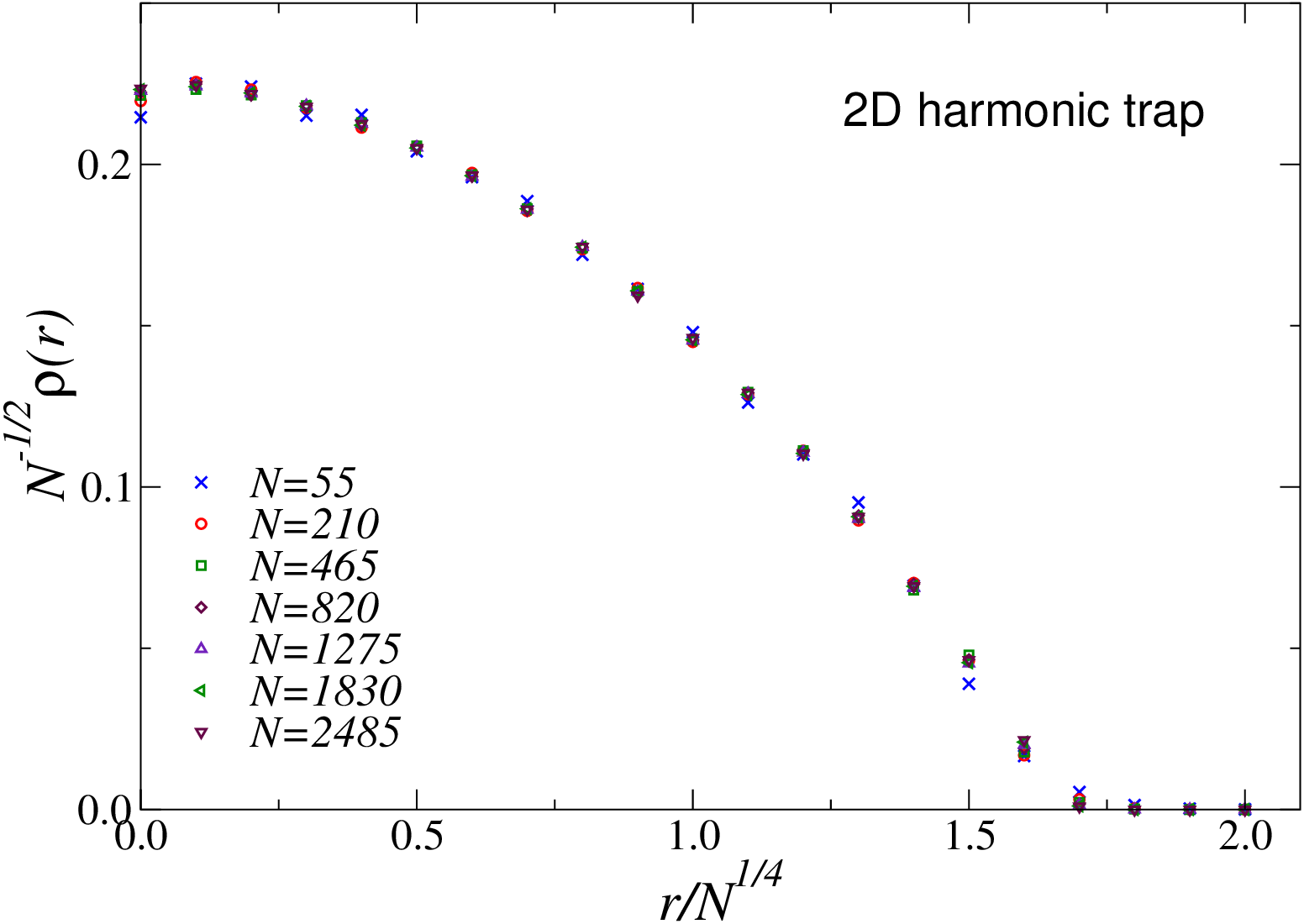}
\includegraphics*[scale=\graphicscale]{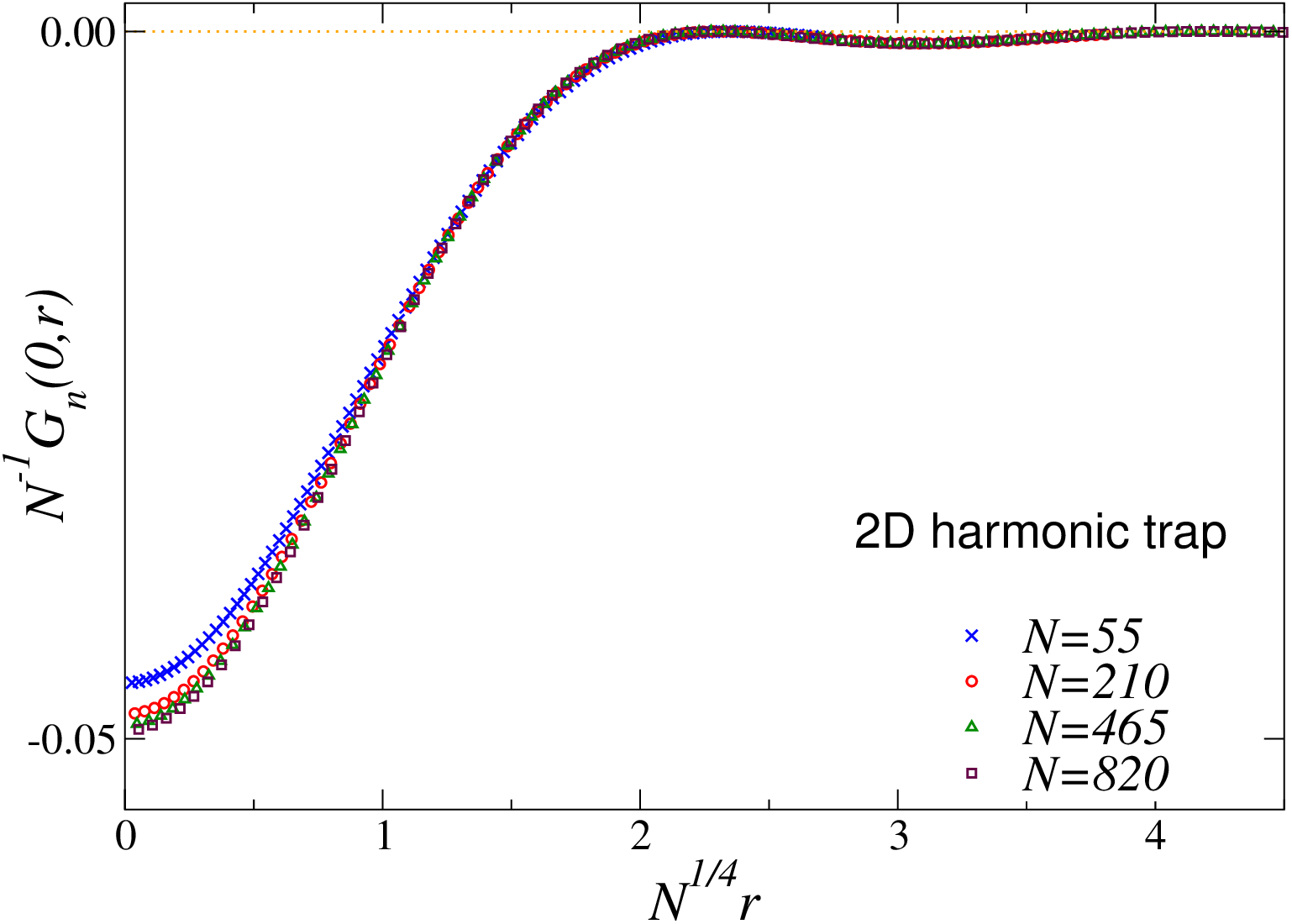}
%=% \vskip-5mm
\caption{(Color online) The particle density and the density
correlator for 2D systems in a harmonic trap: $N^{-1/2}\rho(r)$ vs
$r/N^{1/4}$ (top) and $N^{-1}G_n(0,\vec{x})$ vs $N^{1/4}r$ (bottom)
where $r\equiv|\vec{x}|$ is the distance from the center of the trap.
}
\label{rho2dht} 
\end{figure}

\begin{figure}[tbp]
\includegraphics*[scale=\graphicscale]{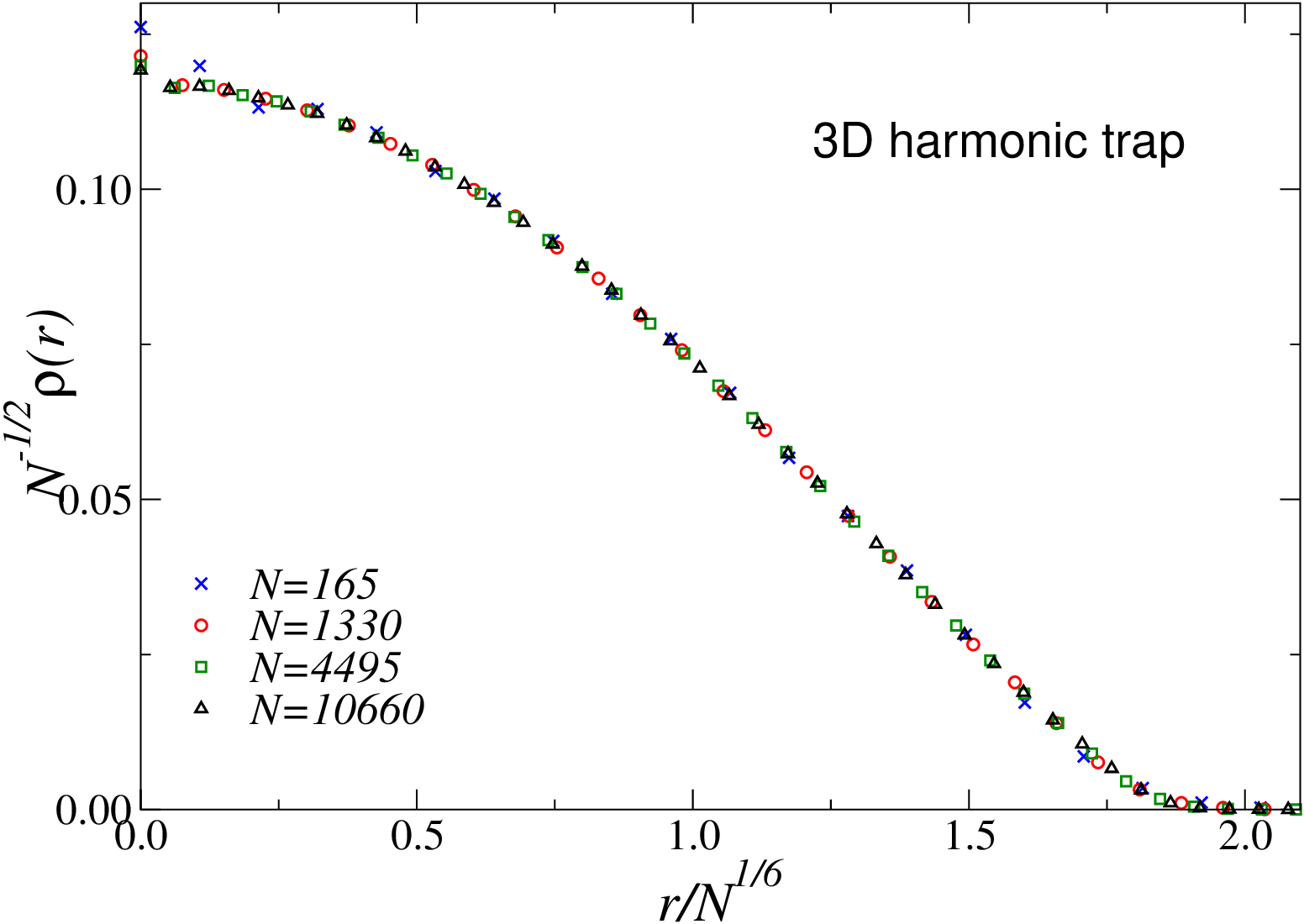}
\includegraphics*[scale=\graphicscale]{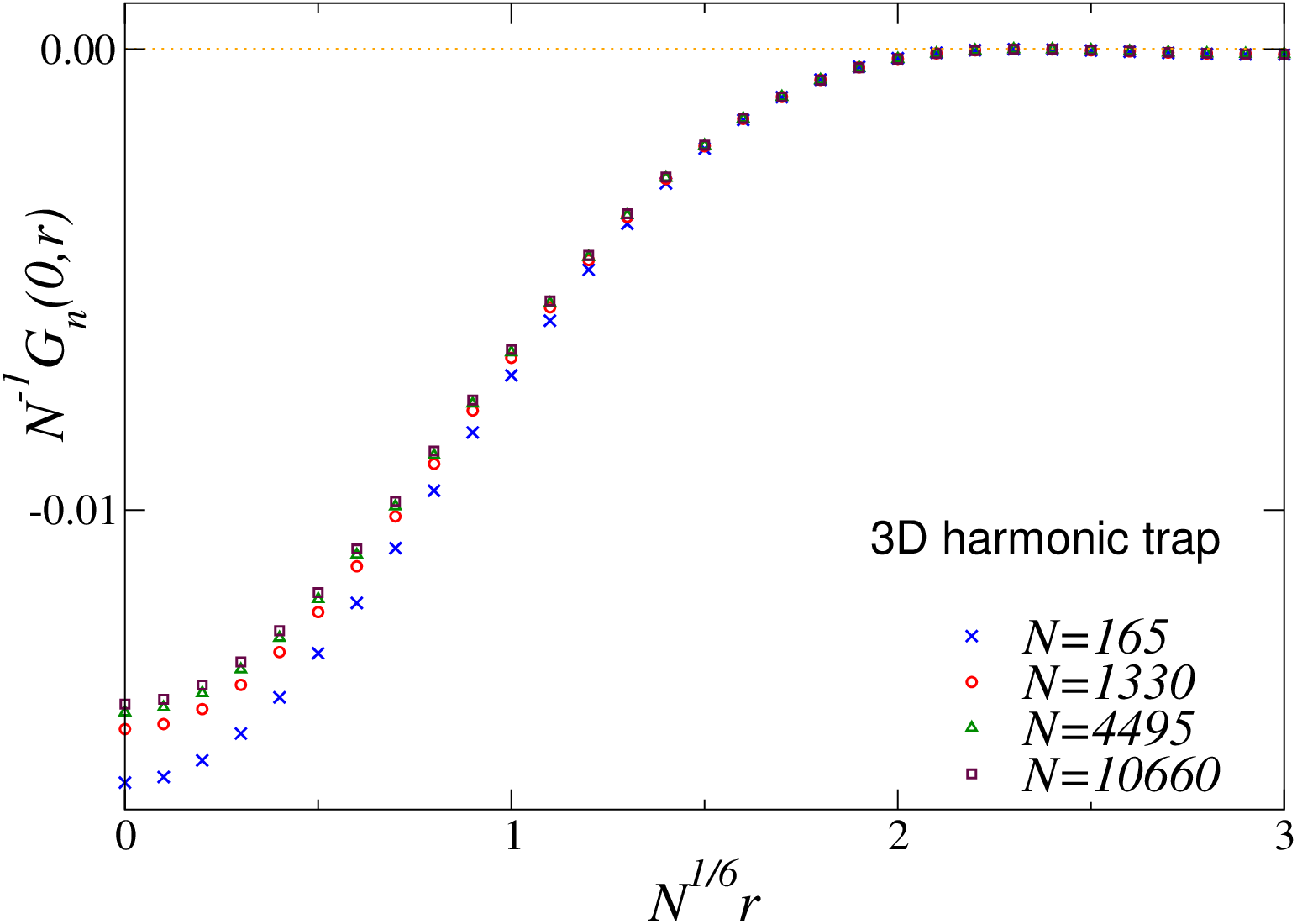}
%=% \vskip-5mm
\caption{(Color online) The particle density and the density
correlator for 3D systems in a harmonic trap: $N^{-1/2}\rho(r)$ vs
$r/N^{1/6}$ (top) and $N^{-1}G_n(0,\vec{x})$ vs $N^{1/6}r$ (bottom)
where $r\equiv|\vec{x}|$ is the distance from the center of the trap.
}
\label{rho3dht} 
\end{figure}

Using the results of Sec.~\ref{genrel}, we can easily obtain results
for the particle density and the density correlator in the presence of
trap. Some data for 2D and 3D systems in a harmonic trap are shown in
Figs.~\ref{rho2dht} and \ref{rho3dht}.  They show the scaling
behavior
\begin{eqnarray}
&&\rho({\bf x}) \approx N^{1/2} R_\rho(r/N^{1/(2d)}),\label{rhox2dht}\\
&&G_n(0,{\bf x}) \approx N R_G(0,rN^{1/(2d)}),\label{grhox2dht}
\end{eqnarray}
where $r\equiv |{\bf x}|$.  Note that, even in dimensions higher than
one, the large-$N$ space rescalings of the particle density and its
connected correlation are different.

\subsection{Half-space entanglement entropies and particle fluctuations}
\label{speng}

\begin{figure}[tbp]
\includegraphics*[scale=\graphicscale]{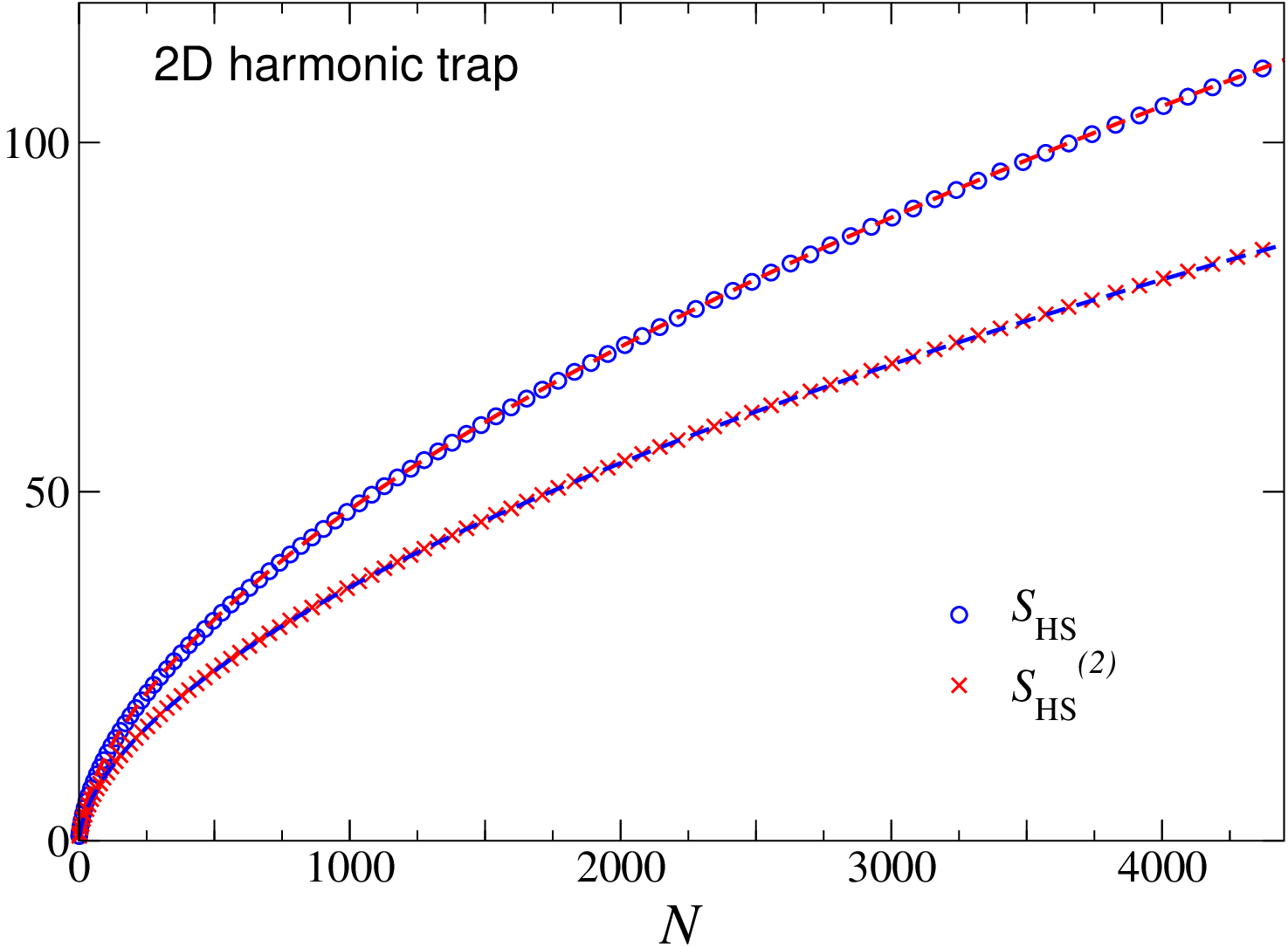}
\includegraphics*[scale=\graphicscale]{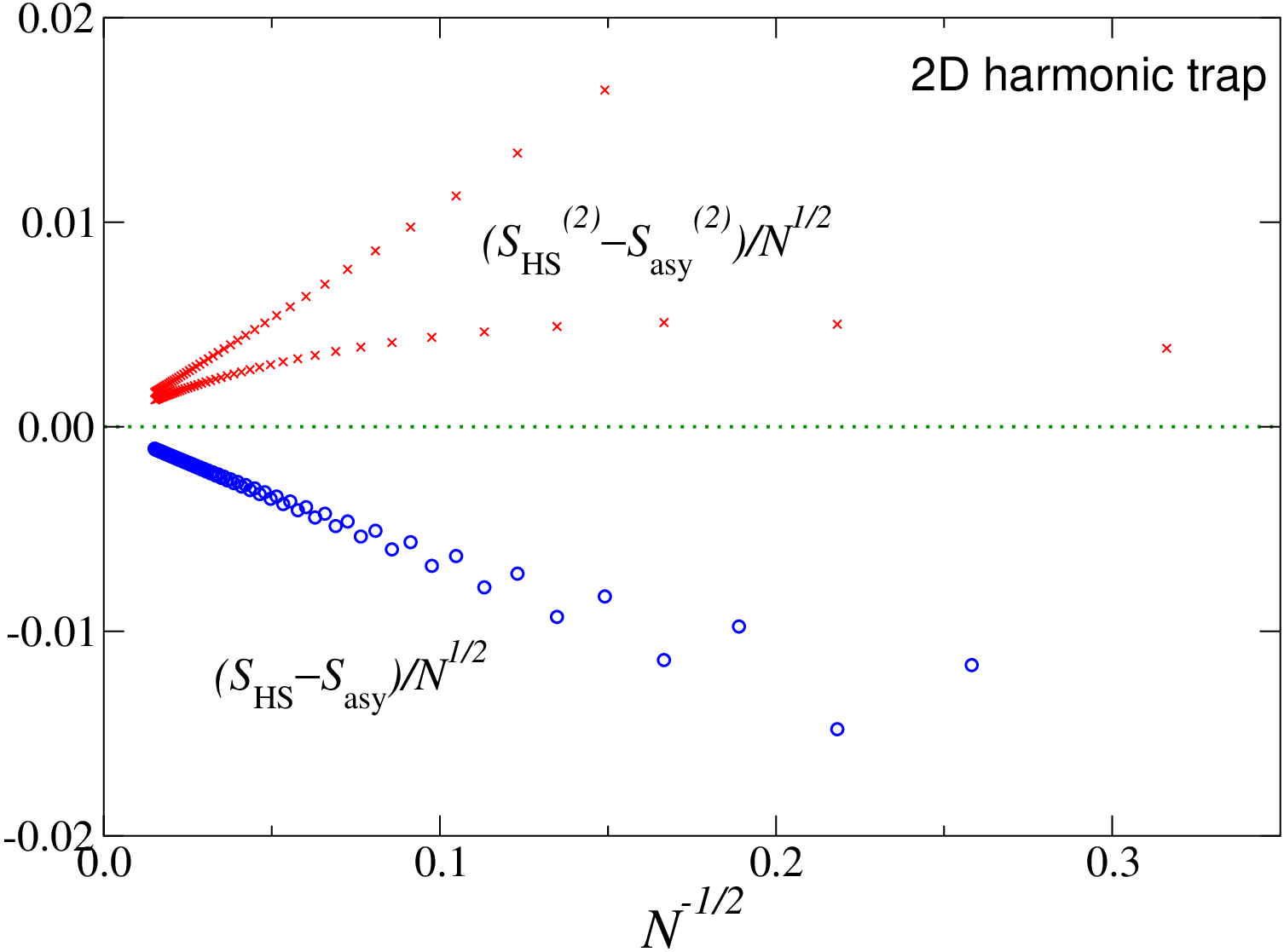}
%=% \vskip-5mm
\caption{(Color online) In the top figure we show the half-space vN
and $\alpha=2$ R\'enyi entanglement entropies of 2D systems with
harmonic trap.  The dashed lines show the predicted asymptotic
behaviors (\ref{sadetrapr}).  In the bottom figure we plot subtracted
entanglement entropies to further check the large-$N$ convergence to
Eq.~(\ref{sadetrapr}), which is clearly demonstrated by the data.  }
\label{s12d}
\end{figure}

In homogeneous systems with periodic and open boundary conditions, the
half-space entaglement entropies of a square $L^2$ system with open
boundary conditions behave as~\cite{CMV-12b}
\begin{equation} 
S_{\rm HS}^{(\alpha)} \approx  c N^{1/2} \ln N,
\quad c = {1+\alpha^{-1}\over 12\pi^{1/2}}. 
\label{shs2dhw}
\end{equation}
The asymptotic large-$N$ behavior of the half-space particle cumulants
and R\'enyi entanglement entropies can be also computed analytically
in the presence of an external harmonic potential.  
For this purpose, we exploit the fact that the corresponding
overlap matrix (\ref{aiodef}) is a block diagonal matrix. Indeed,
relabeling the indeces $n,m$ of the $N\times N$ overlap matrix as
$n_1,...,n_d$, using Eq.~(\ref{prodfunc}), we can write the half-space
overlap matrix as
\begin{equation}
{\mathbb A}_{n_1,...,n_d;m_1,...,m_d} =
\prod_{i=2}^d \delta_{n_im_i} 
 \int_0^\infty dz\, \phi_{n_1}(z) \phi_{m_1}(z)
\label{ahs}
\end{equation}
where $\phi_{n}$ are the 1D eigenfunctions (\ref{1deigf}),
the indeces $n_1,...,n_d$ correspond to the 
lowest $N$ states according to the Eqs.~(\ref{sunei})
and (\ref{Ekphih}).

Let us first consider a 2D system.  We construct the
ground state of a Fermi gases by filling all states with
\begin{equation}
n_1+n_2\le n_e,\qquad n_i,n_e = 1,2,3....\label{nedef}
\end{equation}
The number $N$ of particles is a function of $n_e$, which
asymptotically reads $N=n_e^2/2$.  Since the overlap matrix
(\ref{ahs}) is block diagonal, for any integer $k$ we have
\begin{eqnarray}
&&{\rm Tr}{\mathbb A}_{\rm 2D}[N(n_e)]^k = 
\sum_{n_1=1}^{n_e} {\rm Tr} {\mathbb A}_{\rm 1D}(n_e-n_1)^k , 
\label{adetrap}\\
&&{\mathbb A}_{\rm 1D}(M)_{nm} =  \int_0^\infty dz\, \phi_{n}(z) \phi_{m}(z),
\end{eqnarray}
where ${\mathbb A}_{\rm 1D}(M)$ is the half-space $M\times M$ overlap
matrix of the 1D system.  This also implies analogous exact relations
for all observables which can be constructed by traces of powers of
the overlap matrix or from its eigenvalues, such as the particle
cumulants and the entanglement entropies, cf. Eqs.~(\ref{vny})
and (\ref{snx2n}). Thus,
\begin{eqnarray}
&&S_{\rm HS,2D}^{(\alpha)}[N(n_e)] = 
\sum_{n_1=1}^{n_e} S_{\rm HS,1D}^{(\alpha)}(n_e-n_1) ,
\label{sadetrap}\\
&&V_{\rm HS,2D}^{(m)}[N(n_e)] = 
\sum_{n_1=1}^{n_e} V_{\rm HS,1D}^{(m)}(n_e-n_1) .
\label{vadetrap}
\end{eqnarray}
In order to derive their large-$N$ asymptotic behaviors, we replace the
sums by integrals and use the relation $N=n_e^2/2$, i.e. 
\begin{eqnarray}
&&S_{\rm HS,2D}^{(\alpha)}(N) = 
\int_0^{\sqrt{2N}} dn S_{\rm HS,1D}^{(\alpha)}(\sqrt{2N}-n), 
\label{sadetrapi}\\
&&V_{\rm HS,2D}^{(m)}(N) = 
\int_0^{\sqrt{2N}} V_{\rm HS,1D}^{(m)}(\sqrt{2N}-n).
\label{vadetrapi}
\end{eqnarray}
Then we use the asymptotic formulas for the 1D quantities,
cf.  Eqs.~(\ref{asyt1o2}), (\ref{sasya}), (\ref{v2hstrap}),
(\ref{v2hstraphw}), obtaining
\begin{eqnarray}
&&S_{\rm HS,2D}^{(\alpha)}(N) \approx 
a N^{1/2}\left[ \ln N + a_0 + o(N^0)\right],\label{sadetrapr}\\
&& a = {C_\alpha\over \sqrt{2}} ={1+\alpha^{-1}\over 12\sqrt{2}}
,\qquad 
a_0 = 2 y_\alpha -2 + 7 \ln 2.
\nonumber
\end{eqnarray}
The approximations used to derive this asymptotic behavior from
Eq.~(\ref{sadetrap}) should not affect the leading $O(N^{1/2}\ln N)$
and next-to-leading $O(N^{1/2})$ term, so that the constants $a$ and
$a_0$ should be considered as exact.  This is confirmed by the
analysis of the large-$N$ behavior of numerical data at fixed $N$.  In
Fig.~\ref{s12d} we compare these asymptotic expansions with the data
up to $N\approx 5000$ for the vN and $\alpha=2$ R\'enyi entropy, which
clearly support them.

For the particle cumulants we obtain
\begin{eqnarray}
&&V_{\rm HS,2D}(N) =
v N^{1/2}\left[ \ln N + v_0 + o(N^0)\right],\label{v2adetrapr}\\
&& v = {1\over 2^{3/2} \pi^2} ,\qquad 
v_0 = 2\gamma_E + 7 \ln 2,
\nonumber
\end{eqnarray}
and
\begin{eqnarray}
&&V_{\rm HS,2D}^{(m)}(N) \approx 
\sqrt{2}\nu_m N^{1/2} ,\quad m>2,
\label{vmadetrapr}
\end{eqnarray}
where $\nu_m$ are the constants of the leading large-$N$ behavior in one
dimension, cf. Eq.~(\ref{v2ih}).

The above calculations can be straightforwardly extended to higher
dimensions.  In three dimensions we obtain
\begin{eqnarray}
&&S_{\rm HS,3D}^{(\alpha)}(N) \approx 
{C_\alpha\over \sqrt{6}} 
N^{2/3}  \ln N. \label{sadetrapr3d}
%a_0 = 3 y_\alpha - 9/2 + 10 \ln 2 + \ln 3
\end{eqnarray}

\begin{figure}[tbp]
\includegraphics*[scale=\graphicscale]{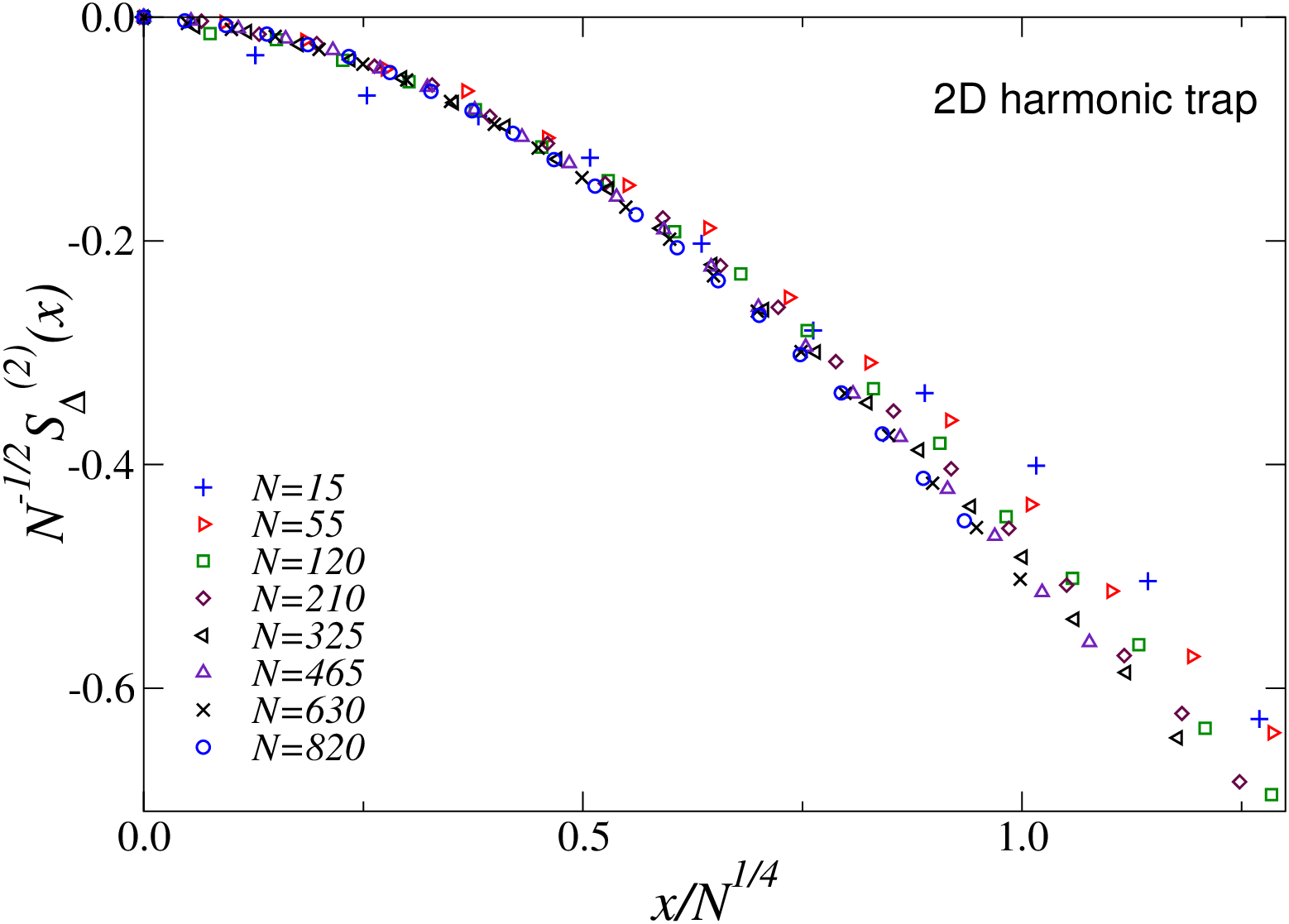}
\includegraphics*[scale=\graphicscale]{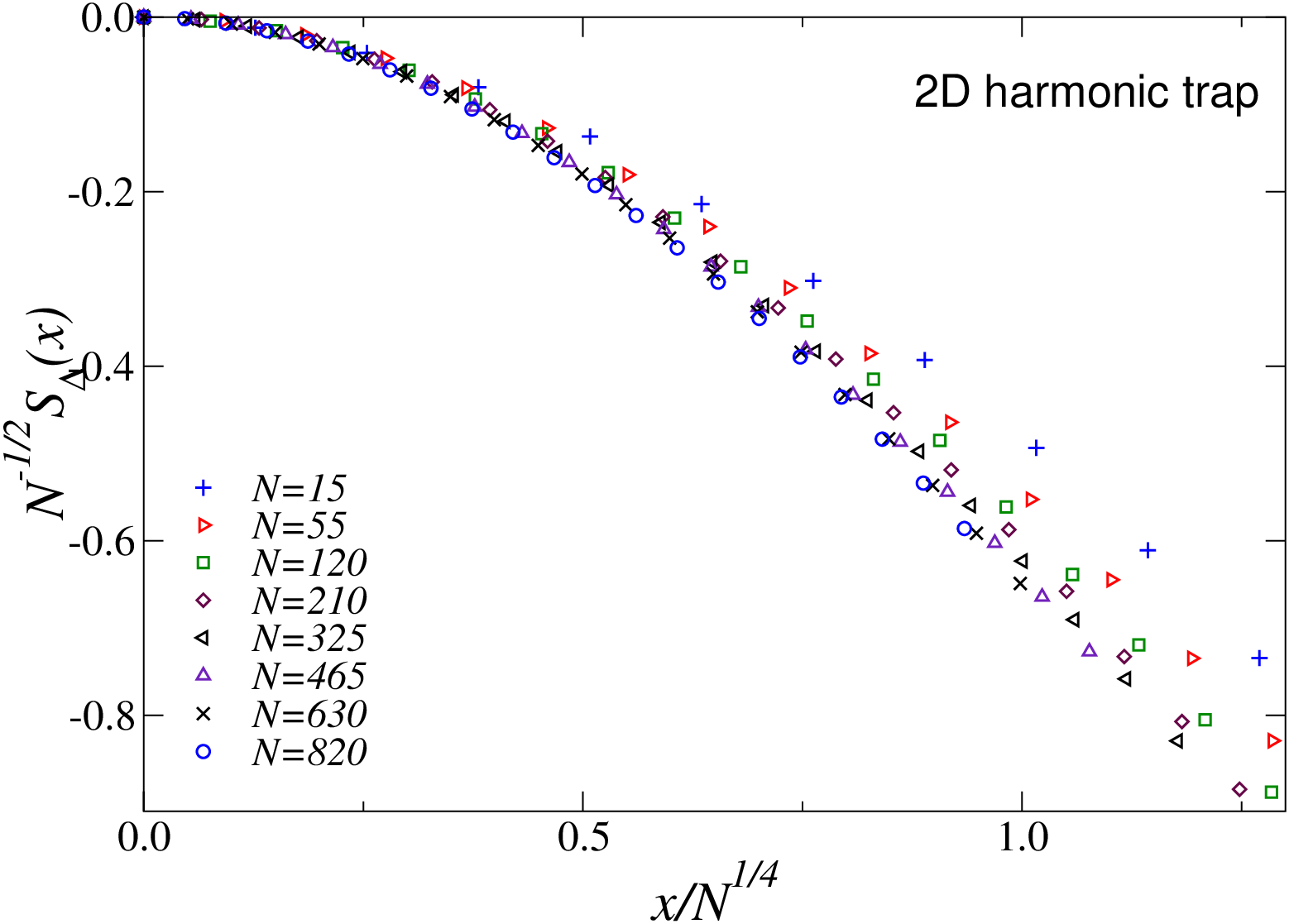}
%=% \vskip-5mm
\caption{(Color online) The space dependence of the $\alpha=2$ R\'enyi
(top) and vN (bottom) entanglement entropies of 2D systems trapped by
a harmonic potential. 
We plot $N^{-1/2}S_\Delta^{(\alpha)}(x)$ vs
$x/N^{1/4}$ for the harmonic trap.  
}
\label{u12dht}
\end{figure}

\begin{figure}[tbp]
\includegraphics*[scale=\graphicscale]{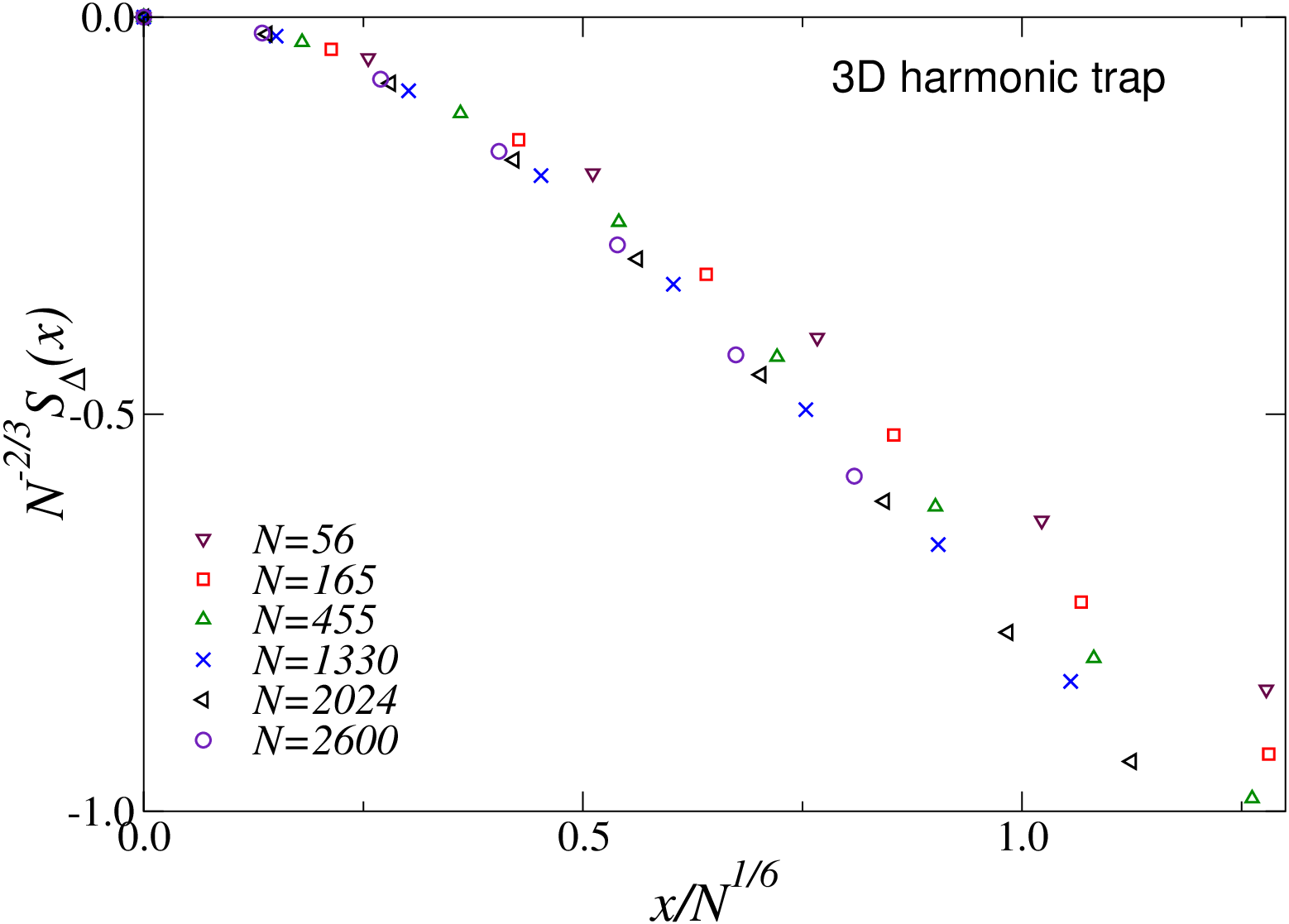}
%=% \vskip-5mm
\caption{(Color online) The space dependence of vN entanglement
entropies of 3D systems trapped by a harmonic potential. We plot
$N^{-2/3}S_\Delta^{(\alpha)}(x)$ vs $x/N^{1/6}$ for the harmonic trap.
}
\label{u12dht3d}
\end{figure}

The above method can be also used to express the entanglement entropies
and particle cumulants of the subsystems $B$ and $S$, defined
at the end of Sec.~\ref{pde2}, in terms of sum of 1D contributions
for the interval $[-\infty,x]$ and $[-x,x]$ respectively.
In particular,  in the case of 
2D  stripes $S$ contained within two parallel lines at distance $x$ from the
center,  the overlap matrix is 
\begin{equation}
{\mathbb A}_{n_1,...,n_d;m_1,...,m_d} =
\prod_{i=2}^d \delta_{n_im_i} 
 \int_{-x}^x dz\, \phi_{n_1}(z) \phi_{m_1}(z),
\label{ahsx}
\end{equation}
which leads to equations analogous to Eqs.~(\ref{adetrap}-\ref{vadetrap}).
Then, using the continuum approximation and the asymptotic large-$N$ behaviors
of the 1D entanglement enetropies of the interval $[-x,x]$,  cf. Eq.~(\ref{smxx}),
we arrive at the asymptotic behavior
\begin{eqnarray}
S_{S, {\rm 2D}}^{(\alpha)}(x) &\approx&  \sqrt{2} C_\alpha
N^{1/2}\Bigl[  \ln N + h_{S^{(\alpha)}}(x/N^{1/4})\Bigr].
 \label{xdeps2bc}
\end{eqnarray}
The coefficient of the leading logarithmic term is just twice that of the
half-space entanglement entropy (\ref{hseb}), 
because the boundary of the stripe is double.  
Analogously, in three dimensions, we obtain
\begin{eqnarray}
S_{S, {\rm 3D}}^{(\alpha)}(x) &\approx&  {\sqrt{2} C_\alpha\over \sqrt{3}}
N^{2/3}\Bigl[  \ln N + h_{S^{(\alpha)}}(x/N^{1/6})\Bigr].
 \label{xdeps2bcd}
\end{eqnarray}
Note that the large-$N$ scaling of the space
variables depends on the spatial dimension $d$.
For a generic $d$, it depends on the scaling variable $x/N^{1/(2d)}$.

The large-$N$  scaling of the space dependence of the entanglement entropies from
the size of the extended space region can be also checked from the
difference $S_\Delta(x)= S_B(x)-S_{\rm HS}$ 
(we recall
that the subsystem $B$ is separated from the rest by a hyperplane at a
distance $x$ from the center of the trap). 
Fig.~\ref{u12dht} shows numerical results for the $\alpha=1$ vN and
the $\alpha=2$ R\'enyi entropyes, which support the large-$N$ scaling
$S_\Delta^{(\alpha)}(x) = N^{1/2} f_\Delta^{(\alpha)}(x/N^{1/4})$,
i.e. the same large-$N$ scaling of the space dependence as in Eq.~(\ref{xdeps2bc}).
Fig.~\ref{u12dht3d} shows the vN
$S_\Delta(x)$ for 3D systems, which appear to scale as
$S_\Delta(x) = N^{2/3} f_\Delta(x/N^{1/6})$.

Analogous results are obtained for the particle variance. In particular,
we find that the ratios of the coefficients of the leading terms
in the asymptotic behaviors of the entanglement entropies and the
particle variance satisfy the universal relation
\begin{equation}
S_A^{(\alpha)}/V_A \approx {(1+\alpha^{-1})\pi^2\over 6} + O(1/\ln N)
\label{savaas}
\end{equation}
for any subsystem $A$ considered and in any dimension.

\section{Conclusions}
\label{conclu}

We investigate the quantum correlations arising in the ground state of
free fermion gases trapped by an external space-dependent harmonic
potential, $V \propto x^2/l^2$ where $l$ is the trap size, in one, two
and three dimensions.  We consider systems of $N$ particles, and focus
on the large-$N$ scaling behaviors of the quantum correlations, as
inferred by the expectation values of product of local operators and
bipartite entanglement entropies which quantify the nontrivial
entanglement connections between different parts of extended quantum
systems.  In particular, we study the relations between the
entanglement entropies and the cumulants of the particle distribution
within the same extended subsystem, which can be obtained by
integration of the particle-density correlations.

Our results for the large-$N$ behaviors of the particle density
$\rho(x)$, the two-point particle correlation $C(x,y)$ and the
connected density-density correlation $G_n(x,y)$, can be summarized by
the following scaling equations:
\begin{eqnarray}
&&\rho(r) \approx N^\theta \xi^{-d} R_\rho(N^{(\theta-1)/d}r/\xi),
\quad r\equiv|{\bf x}|, 
\label{rhoxlnb}
\end{eqnarray}
and
\begin{eqnarray}
&&C({\bf x}_1,{\bf x}_2) \approx N^{\theta} \xi^{-d}
R_C(N^{\theta/d} {\bf x}_1/\xi,N^{\theta/d} {\bf x_2}/\xi),\;\;
\label{gxlnb}\\
&&G_n({\bf x}_1,{\bf x}_2) \approx N^{2\theta} \xi^{-2d}
R_G(N^{\theta/d} {\bf x}_1/\xi,N^{\theta/d} {\bf x}_2/\xi),\qquad
\label{gnxlnb}
\end{eqnarray}
for ${\bf x}_1\ne {\bf x}_2$, where $d$ is the spatial dimension of
the system, $\xi\equiv l^{\theta}$ is the (oscillator) length scale
induced by the trap, and $\theta=1/2$ is the trap exponent for the
harmonic potential.  The above large-$N$ behaviors are
expected to also hold for higher power laws of the external potential,
i.e. $V(r) \propto (r/l)^{p}$, by replacing the corresponding value
of the trap exponent, i.e.  $\theta\equiv p/(p+2)$. In the limit
$p\to\infty$, corresponding to hard-wall trap, the scaling laws of
homogeneous systems are recovered by setting $\theta=1$.
 
We compute and analyze the asymptotic large-$N$ behaviors
of the particle cumulants and entanglement entropies of extended
spatial regions. Our main results are:
 
(i) The half-space R\'enyi entanglement entropies behave as
\begin{eqnarray}
S_{{\rm HS}}^{(\alpha)} = {1+\alpha^{-1}\over 2} c_l N^{(d-1)/d} \left[
\ln N + c_0 + o(1) \right],
\label{hseb}
\end{eqnarray}
which includes the vN entanglement entropy when $\alpha\to 1$.  In
1D systems, the constant of the logarithmic term is equal
to that of the homogeneous system, i.e. $c_l=1/6$, which is related to
the central charge $c=1$ of the corresponding conformal field
theory~\cite{CC-04,CMV-11}.  We also determine the subleading
constant $c_0$, cf. Eqs.~(\ref{asyt1o2}-\ref{sasya}).  We also obtain
the constants $c_l$ and $c_0$ in higher dimensions,
cf. Eqs.~(\ref{sadetrapr}) and (\ref{sadetrapr3d}); in particular we
find $c_l=1/(6\sqrt{2})$ and $c_l=1/(6\sqrt{6})$ for the leading
logarithmic term in two and three dimensions respectively.

(ii) We compute the asymptotic large-$N$ behavior of the half-space
particle cumulants.  Only even cumulants are nonzero, because
half-space odd cumulants vanish by symmetry.  We obtain
\begin{eqnarray}
&&V_{{\rm HS}} = v_l N^{(d-1)/d} \left[ \ln N + v_0 + o(1) \right]
\label{v2co}\\
&&V_{{\rm HS}}^{(2k)} \approx w_{2k} N^{(d-1)/d},\quad k\ge 2, 
\label{v2kco}
\end{eqnarray}
In one dimension, see  Eqs.~(\ref{v2hstrap}) and (\ref{v2ih}), the
constants of the leading terms $v_l$ and $w_{2k}$ turn out to be equal
to those of the homogeneous system with open boundary conditions (hard
walls), which were already computed in Refs.~\cite{CMV-12l}
(see also Refs.~\cite{ELR-06,SRFKLL-12}). In particular,
$v_l=1/(2\pi^2)$ and $v_0$ is reported in Eq.~(\ref{v2hstrap}).  
The constants $v_l$ and $v_0$ are also  evaluated
in higher dimensions, cf. Eqs.~(\ref{v2adetrapr}) and
(\ref{vmadetrapr}). Only the particle variance presents
the leading logarithmic term, like 
homogeneous systems.
We find that, in any dimension and for any subsystem $A$,
the ratio of the coefficients of the leading terms in the
entanglement entropies and particle variance satisfies the relation
\begin{equation}
{c_l\over v_l}={\pi^2\over 3}
\label{clvl}
\end{equation}

(iii) We also consider spatial bipartitions with different geometries,
in particular the entanglement entropy of a {\em stripe} $S$ around the
center of the trap with the boundaries at a distance $x$
(in 1D $S=[-x,x]$), defined in
Sec.~\ref{pde2}, and studied its space dependence.  We find the
general behavior
\begin{eqnarray}
S_S^{(\alpha)}(x) &\approx&  {1+\alpha^{-1}\over 2} 2 c_l
N^{(d-1)/d}\Bigl[
 \ln N + \label{xdeps2b}\\
&&+f_{S^{(\alpha)}}(N^{(\theta-1)/d}x/\xi)\Bigr],\nonumber
\end{eqnarray}
where $c_l$ is the same constant appearing in Eq.~(\ref{hseb}).  
The
coefficient of the leading logarithmic term is just twice that of the
half-space entanglement entropy (\ref{hseb}), in any dimension,
essentially because the boundary of the stripe is double.  
A detailed analysis of the 1D case is reported
in Sec.~\ref{1dseb}. The particle variance shows an analogous
behavior, i.e.
\begin{eqnarray}
V \approx  2 v_l
N^{(d-1)/d}\left[
 \ln N + 
+f_{V}(N^{(\theta-1)/d}x/\xi)\right],
\label{va1dco}
\end{eqnarray}
where $v_l$ is the same constant appearing in Eq.~(\ref{v2co}).
Note that the large-$N$ scaling of the space dependence of the
entanglement entropies and particle variance is analogous to that of
the particle density, while it differs from that of the particle correlation
$C(x,y)$ and the connected density correlation $G_n(x,y)$,
cf. Eqs.~(\ref{gxlnb}) and (\ref{gnxlnb}).

The above results (i), (ii) and (iii) are consistent with the known
asymptotic behaviors of homogeneous Fermi gases with open boundary
conditions~\cite{CMV-11a,CMV-12b,CMV-12l}, obtainable by setting
$\theta=p/(p+2)\to 1$.

The large-$N$ asymptotic behaviors are rapidly approached with increasing
the number of particles. For example, in one dimension the behavior of
$O(10^2)$, of even less, particles is already well approximated by the
asymptotic behaviors.

Finally, a few comments are in order concerning the relations between
particle cumulants and entanglement entropies of an extended subsystems
$A$.  For noninteracting fermions, one can write down a formal
expansion of the entanglement entropies of bipartitions in terms of
the even cumulants~\cite{KRS-06,KL-09,SRL-10,SRFKL-11}, such as
\begin{eqnarray}
&&S_A = {\pi^2\over 3} V_A
+ {\pi^4\over 45} V^{(4)}_A
+ {2\pi^6\over 945} V^{(6)}_A + ...,
\label{s1vn} \\
&&S^{(2)}_A = {\pi^2\over 4}
V_A - {\pi^4\over 192} V^{(4)}_A + {\pi^6\over 23040} V^{(6)}_A
+ ...
\label{s2vn}
\end{eqnarray}
In homogeneous noninteracting fermion gases with $N$ particles in a
finite volume of any dimension $d$, the above expansions gets
effectively truncated in the large-$N$ limit~\cite{CMV-12l} because
the high cumulants $V^{(m)}_A$ with $m>2$ are all suppressed
relatively to the particle variance $V_A$.  The leading $N^{(d-1)/d}
\ln N$ asymptotic behavior of $S^{(\alpha)}_A$ in Eqs.~(\ref{s1vn})
and (\ref{s2vn}) arises from $V_A$ only, because the leading order of
each cumulant $V^{(k)}$ with $k>2$ vanishes for any subsystem $A$
(including disjoint ones) in any dimension.  This implies the general
asymptotic relation
\begin{equation}
S^{(\alpha)}_A \approx  
{(1+\alpha^{-1})\pi^2\over 6} V_A . \label{anyd}
\end{equation}
Our results for Fermi gases trapped by a harmonic potential show an
analogous scenario: the asymptotic relation (\ref{anyd}) holds as
well, in any dimension, significantly extending its validity.

We mention that the close relation between entanglement entropy and
variance in noninteracting Fermi gas is also found in off-equilibrium
phenomena after local quantum quenches~\cite{KL-09,hgf-09,SRFKLL-12},
and in some dynamics regime of the off-equilibrium expansion of Fermi
gases from a trap~\cite{VVV}.

The situation is more involved for interacting systems.  In
systems with localized interactions arising from
impurities~\cite{CMV-12a}, all the cumulants $V^{(2k)}$ contribute to
the asymptotic large-$N$ behavior of the entanglement entropies in the
expansion $S^{(\alpha)}_A= \sum_{k=1}^\infty s^{(\alpha)}_k
V^{(2k)}_A$, although the expansion turns out to be rapidly
converging~\cite{CMV-12l}.  The conservation of a global charge, and
in particular the particle number, is crucial for the connections
between bipartite entanglement and particle fluctuations. For
interacting systems not conserving the particle number, the
entanglement should be related to the more fundamental energy
transport.

\acknowledgements
%{\it Acknowledgements}.  
I thank Pasquale Calabrese and Mihail Mintchev for many useful
discussions within common research projects.

\appendix

\section{Asymptotic behavior of the 1D half-space entanglement 
entropies}
\label{bentBH}

We consider a 1D lattice model of spinless fermions in
the presence of an external space-dependent potential
\begin{eqnarray}
H_f = =\sum_{\langle ij \rangle} c_{i}^\dagger h_{ij} c_{j} 
\label{fermmod}
\end{eqnarray}
where $c_i$ is a spinless fermion operator,
and
\begin{eqnarray}
h_{ij} = J \left(\delta_{ij} - {1\over 2} \delta_{i,j-1} 
- {1\over 2} \delta_{i,j+1}
\right)
+ V(x_i) \delta_{ij}. 
\label{hijdef}
\end{eqnarray}
We consider a power-law spatial dependence for the trapping
potential,
\begin{equation}
V(r) = {1\over p} v^p r^p, 
\label{potential}
\end{equation}
where $r$ is the distance from the center of the trap, $v$ is a
positive constant and $p$ an even integer number.  The trap size $l$
is defined as $l\equiv J^{1/p}/v$.  In the following we set $J=1$.

In one dimension this lattice free-fermion model can be exactly mapped into the
hard-core (HC) limit of the Bose-Hubbard (BH) model, see, e.g.,
Ref.~\cite{Sachdev-book},
\begin{eqnarray}
{\cal H}_{\rm BH}& =& {J\over 2}
\sum_{\langle ij\rangle} (b_j-b_i)^\dagger (b_j-b_i)\label{bhmN}\\
&&+ {U\over2} \sum_i n_i(n_i-1) + \sum_i V(r_i) n_i ,
\nonumber
\end{eqnarray}
where $\langle ij\rangle$ is the set of nearest-neighbor sites, $b_i$
are bosonic operators, $n_i\equiv b_i^\dagger b_i$ is the particle
density operator, and $N = \langle \sum_i n_i \rangle$ is the particle
number.  The HC limit $U\to\infty$ of the BH model implies that the
particle number $n_i$ per site is restricted to the values $n_i=0,1$.
As a consequence of their exact mapping, the HC BH and lattice
free spinless fermions share the same particle density and
density-density correlation, particle distribution of extended
space regions,
and also entanglement entropies of
connected spatial bipartitions~\cite{rev-cc}.

The large-$l$ limit keeping $N$ fixed differs from that performed at
fixed chemical potential $\mu$, i.e., considering the BH Hamiltonian
\begin{eqnarray}
{\cal H}_{\rm \mu} =  {\cal H}_{\rm BH}  + (\mu-1) \sum_i n_i .
\label{bhm}
\end{eqnarray}
Indeed, the large trap-size limit, keeping $\mu$ fixed, implies an
increase of the particle number so that
\begin{equation}
N/l^d = \tilde{\rho}(\mu)
\label{rhonld}
\end{equation}
asymptotically, where $d$ is the spatial dimension and
$\tilde{\rho}(\mu)$ is a finite function of $\mu$.  This {\em
thermodynamic} limit is usually considered when quantum transitions
are studied in confined particle systems.  In the absence of the trap,
the 1D HC BH model has three phases: the empty state for $\mu>1$ with
$\langle n_i\rangle=0$, which may be seen as a particular $n=0$ Mott
phase, a gapless superfluid phase for $|\mu|< 1$, and a $n=1$ Mott
phase for $\mu<-1$.  See, e.g., Ref.~\cite{Sachdev-book}.

We consider chains with even sites $L$ and open boundary conditions,
and a trap of size $l$ centered between the middle sites of the chain.
We divide the chain in two connected parts of length $l_A$ and
$L-l_A$, and consider their R\'enyi entropies
\begin{equation}
{\cal S}^{(\alpha)}(l_A;L) = {\cal S}^{(\alpha)}(L-l_A;L) = 
{1\over 1-\alpha} \ln {\rm Tr} \rho_A^\alpha
\label{renyientropies}
\end{equation}
where $\rho_A$ is the reduced density matrix of one of the two
subsystems.  Let us consider the half-space entanglement
\begin{equation}
{\cal S}_{{\rm HS}}^{(\alpha)}\equiv S^{(\alpha)}(L/2,L)
\label{hsedef}
\end{equation}
Its large-$L$ behavior can be written as
\begin{eqnarray}
&& {\cal S}_{{\rm HS}}^{(\alpha)} = C_\alpha\left[\ln L
+ e_\alpha + O(L^{-1/\alpha}) \right],\qquad \label{pinfbeh1o2}\\
&&C_{\alpha}={1+\alpha^{-1}\over 12},\label{caaa}\\
&&e_\alpha = \ln\sqrt{1-\mu^2} + \ln (4/\pi) + y_\alpha,\label{yaca}\\
&& y_\alpha =
\int_0^\infty {dt\over t} \Bigl[{6\over 1-\alpha^{-2}}
\left({1\over \alpha\sinh t/\alpha} - {1\over\sinh t}\right)\times
\nonumber \\
&&\qquad\quad\times{1\over\sinh t}- e^{-2t}\Bigr]
\label{yalpha}
\end{eqnarray} 
where $y_\alpha$ is given in Eq.~(\ref{yalpha}).
This equation has been  obtained
using the results of Refs.~\cite{JK-04,CC-04,IJ-08,CC-10}.

In a system confined by a trap of size $l$, we have~\cite{CV-10-e}
\begin{eqnarray}
&& {\cal S}_{{\rm HS}}^{(\alpha)} \equiv  {\rm Lim}_{L\to\infty}
S^{(\alpha)}(L/2;L) = \label{shsdef}\\
&&=C_\alpha\left[
\ln \xi_e + e_\alpha + O(\xi_e^{-1/\alpha})\right]
 \nonumber 
\end{eqnarray}
where 
\begin{equation}
\xi_e = a_e(\mu) l
\label{xie}
\end{equation}
is the entanglement length, which also enters the asymptotic formula
of the energy difference of the two lowest states
\begin{equation}
\Delta = {\pi\sqrt{1-\mu^2}\over \xi_e} t(\phi),\qquad
t(\phi) = 1/2 - |\phi-1/2|, 
\label{gapXX}
\end{equation}
where $\phi$ is phase-like variable, $0\le \phi <1$, which
parametrizes the modulations of the amplitude due to the periodic
asymptotic occurrence of level crossings in the large-$l$
limit~\cite{CV-10-bh}.

We now consider a HC BH system of $N$ particles in a trap (centered in
the middle of the chain), and we want to study the bipartite
entanglement entropy, and in particular the half-lattice entanglement
entropy, in the limit of large trap size (after taking the infinite
chain limit $L\to\infty$) as a function of $N$.  More precisely, we
consider
\begin{eqnarray}
&&S^{(\alpha)}_{\rm HS}(l,N) = {\rm Lim}_{L\to\infty}
S^{(\alpha)}(L/2,L;l,N) ,\quad \label{sal}\\
&&S^{(\alpha)}_{\rm HS}(N) \equiv {\rm Lim}_{l\to\infty} S^{(\alpha)}(l,N),
\label{sali}
\end{eqnarray}
which is a finite function of $N$.  Since $N$ is kept fixed while
$l\to\infty$, this corresponds to the dilute regime $\mu\lesssim 1$.
Let us define $\delta=1-\mu$, thus $\delta\to 0^+$ corresponds to the
dilution limit.

We want to derive asymptotic large-$N$ behavior of $S_{\rm HS}(N)$.
For this purpose, we need: (i) the dependence of $a_e$ on $\mu$ when
$\mu\to 1^-$; (ii) the relation between the number of particles and
the chemical potential.

The dependence of $a_e$ on $\mu$ can be inferred from the behavior of
the gap for $\delta\to 0$ (see Sec. IV B of Ref.~\cite{CV-10-bh}), by
matching Eq.~(\ref{gapXX}) with its asymptotic behavior for $\delta\to
0$
\begin{equation}
\Delta \approx c_p \delta^{(p-2)/(2p)} {t(\phi)\over l} 
\label{deltaasy}
\end{equation}
where
\begin{equation}
c_p(\delta) = \Biggl\{
\begin{array}{cl}
1 & \quad {\rm for} \quad p = 2\\

{2\sqrt{\pi}\Gamma(3/4)\over\Gamma(1/4)} & \quad {\rm for} \quad p=4 \\

{\pi\over \sqrt{2}} & \quad {\rm for} \quad p\to\infty \\
\end{array}
\label{cp}
\end{equation}

In 1D particle systems, the {\em thermodynamic} limit at fixed $\mu$
corresponds to $N,l\to\infty$ keeping the ratio $N/l$ fixed.  Indeed,
we have
\begin{equation}
N \equiv \langle \sum_i b_i^\dagger b_i  
\rangle = \tilde{\rho}(\mu) l + O(1)\label{cmudef}
\end{equation}
The function $\tilde{\rho}(\mu)$ can be computed in the HC limit.  The
particle density in the large-$l$ limit turns out to approach its
local density approximation (LDA), with corrections that are
suppressed by powers of the trap size and present a nontrivial TSS
behaviour.  Within the LDA, the particle density at the spatial
coordinate $x$ equals the particle density of the homogeneous system
at the effective chemical potential
\begin{equation}
\mu_{\rm eff}(x) \equiv \mu + {1\over p} \left({x\over l}\right)^p.
\label{mueff}
\end{equation}
The LDA of the particle density reads
$\langle n_x \rangle_{\rm lda} \equiv \rho_{\rm lda}(x/l)$, where
\begin{equation}
\rho_{\rm lda}(x/l) = 
\kern-10pt \quad\left\{
\begin{array}{l@{\ \ }l@{\ \ }l}
0 & {\rm for} & \mu_{\rm eff}(x) > 1, \\
(1/\pi)\arccos\mu_{\rm eff}(x) &
    {\rm for} & -1 \le \mu_{\rm eff}(x) \le 1, \\
1 & {\rm for} & \mu_{\rm eff}(x) < -1. \\
\end{array} \right.
\label{nxlda}
\end{equation}
Asymptotically, the total particle number is obtained by integrating 
the LDA of the particle density $\rho_{\rm lda}$, obtaining
\begin{eqnarray}
\tilde{\rho}(\mu) = 2  \int_0^\infty \rho_{\rm lda}(y) \,{\rm d} y.
\label{rhomu}
\end{eqnarray}
In the low-density regime, $\delta \equiv 1- \mu\to 0$,
\begin{equation}
\tilde{\rho}(\mu) = r_p \delta^{(p+2)/(2p)} [1 + O(\delta)],
\label{tilderho}
\end{equation}
with
\begin{equation}
r_p = \Biggl\{
\begin{array}{cl}
1 & \quad {\rm for} \quad p = 2 \\

{2 \Gamma(1/4)\over  3 \sqrt{\pi} \Gamma(3/4)} & \quad {\rm for} \quad p=4 \\

{2\sqrt{2}\over \pi} & \quad {\rm for} \quad p\to\infty \\
\end{array}
\label{rp}
\end{equation}

Using these results we obtain the entanglement entropy
(\ref{shsdef}) in terms of $N$:
\begin{eqnarray}
S_{\rm HS}^{(\alpha)}(N)=
 C_\alpha\left[\ln N + \ln {4(p+2)\over p} + y_\alpha
+ o(N^0)\right]. \nonumber \\  \label{snalphab}
\end{eqnarray}
The above formula can be compared with large-$L$ and then large-$l$
extrapolations of numerical results by exact diagonalization of chains
of size $L$ and traps of size $l$ centered between the middle sites.
In particular, the $N$-dependent limit (\ref{snalphab}) is
approached with $O(l^{-2\theta})$ corrections, where $\theta=p/(p+2)$
is the trap exponent, conferming general theoretical arguments on the
large-$l$ corrections~\cite{CV-10-bhn}.  

The large trap-size limit keeping the particle number $N$ fixed
corresponds to the dilute limit of continuum models in the presence of
the trapping potential. As argued in Ref.~\cite{CV-10-bhn}, it also
represents the asymptotic large-$N$ behavior for finite on-site
couplings $U>0$ within the BH model, or finite-strength models of
bosonic gases with short-range interactions.

\end{document}